\documentclass[
acmsmall,
authorversion=true
]{acmart}

\acmJournal{TOSEM}

\usepackage{pifont}
\usepackage{paralist}
\usepackage{algorithmic}
\usepackage{graphicx}
\usepackage{textcomp}
\usepackage{xcolor}
\usepackage{array,booktabs}
\usepackage{multirow}
\usepackage[linesnumbered,ruled]{algorithm2e}
\usepackage{xspace}
\usepackage{pgfplotstable}
\usepackage{paralist}
\usepackage[detect-weight=true, detect-family=true]{siunitx}
\DeclareSIUnit{\pp}{\textup{pp}}
\usepackage{datatool}

\usepackage{subfigure}
\usepackage{tikz}
\usetikzlibrary{shapes.geometric,arrows}
\usetikzlibrary{positioning,fit,calc}
\usetikzlibrary{arrows.meta}
\tikzstyle{arrow}=[draw] 
\usetikzlibrary{arrows}
\usepackage{subfigure}
\usepackage{pgfplotstable}
\pgfplotsset{compat=1.18}
\usepgfplotslibrary{statistics}
\usepackage{makecell}
\usepackage[normalem]{ulem}

\newcommand{\approach}{LACQUER\xspace}
\newcommand{\myblock}[1]{\tikz[baseline=(char.base)]{
		\node[shape=circle, draw, fill=gray, opacity=.2, text opacity=1,inner sep=0.5pt] (char) {\vphantom{1g}#1};}}

\usepackage{hyperref}
\begin{document}

\title{Learning-Based Relaxation of Completeness Requirements for Data Entry Forms}

\author{Hichem Belgacem}
\orcid{0000-0002-0521-2905}
\affiliation{\institution{University of Luxembourg}
  \country{Luxembourg}
}
\email{hichem.belgacem@uni.lu}

\author{Xiaochen Li}
\orcid{0000-0002-5068-1938}
\affiliation{\institution{Dalian University of Technology}
  \country{China}
}  
\email{xiaochen.li@dlut.edu.cn}
\authornote{Part of this work was done while the author was affiliated with
  the University of Luxembourg, Luxembourg.}

\author{Domenico Bianculli}
\orcid{0000-0002-4854-685X}
\affiliation{\institution{University of Luxembourg}
  \country{Luxembourg}
}
\email{domenico.bianculli@uni.lu}

\author{Lionel Briand}
\orcid{0000-0002-1393-1010}
\affiliation{\institution{University of Luxembourg}
  \country{Luxembourg}
}
\affiliation{\institution{University of Ottawa}
  \country{Canada}
}  
\email{lionel.briand@uni.lu}

\setcopyright{acmlicensed}
\acmJournal{TOSEM}
\acmYear{2023} \acmVolume{1} \acmNumber{1} \acmArticle{1} \acmMonth{1} \acmPrice{}\acmDOI{10.1145/3635708}

\begin{abstract}

Data entry forms use completeness requirements to specify the fields that are required or optional to fill
for collecting necessary information from different types of users.
However, because of the evolving nature of software, 
some required fields may not be applicable for certain types of users
anymore. Nevertheless, they may still be incorrectly marked as
required in the form; we call such fields obsolete required fields.
Since obsolete required fields usually have ``not-null'' validation checks before submitting the form, 
users have to enter meaningless values in such fields in order to
complete the form submission.
These meaningless values threaten the quality of the filled data,
and could negatively affect stakeholders or learning-based tools that use the data. 
To avoid users filling meaningless values,
existing techniques usually rely on manually written rules to identify the obsolete required fields and relax their completeness requirements.
However, these techniques are ineffective and costly.

In this paper, we propose \approach, a learning-based automated approach 
for relaxing the completeness requirements of data entry forms. 
\approach builds Bayesian Network models to automatically learn conditions under which users had to fill
meaningless values. 
To improve its learning ability, \approach identifies the cases where a required field is only
applicable for a small group of users, and uses SMOTE, an oversampling technique, to generate more instances on such fields for 
effectively mining dependencies on them.
During the data entry session, \approach predicts the completeness requirement of a target based on the already filled fields 
and their conditional dependencies in the trained model.

Our experimental results show that \approach can accurately relax the completeness requirements of 
required fields in data entry forms with precision values ranging between 0.76 and 0.90 on different datasets.
\approach can prevent users from filling 20\% to 64\% of meaningless values, with negative predictive values (i.e., the ability to correctly predict a field as ``optional'')
between 0.72 and 0.91.
Furthermore,  \approach is efficient; it takes at most
\SI{839}{\milli\s} to predict the completeness requirement of an instance.

 \end{abstract}



\begin{filecontents*}{data/rq-applicability.csv}

Dataset, Algorithm,Prec-S1,Rec-S1,NPV-S1,Spec-S1,Prec-R1,Rec-R1,NPV-R1,Spec-R1,Train,P-Avg, P-Range
    , Ripper,0.63,0.79,0.17,0.20,0.69,0.83,0.25,0.16,349.29,0.18,0.18--0.19
NCBI,ARM,0.75,0.98,0.81,0.16,0.82,0.86,0.39,0.28,11.98,5.06,3--12
    ,LACQUER,0.76,0.98,0.91,0.20,0.84,0.98,0.76,0.37,145.98,75.83,33--144
    ,Ripper,0.66,0.73,0.60,0.29,0.58,0.62,0.84,0.56,240.37, 0.24,0.15--0.54
PEIS,ARM,0.72,0.80,0.24,0.24,0.72,0.80,0.25,0.25,153.78,1.59,2--20
    ,LACQUER,0.88,0.98,0.72,0.62,0.90,0.97,0.75,0.64,1210.70,307,179--839
\end{filecontents*}
\DTLloaddb{effectivenss}{data/rq-applicability.csv}

\DTLgetvalue{\LACQUERSeqSpecNcbi}{effectivenss}{3}{\dtlcolumnindex{effectivenss}{Spec-S1}}
\DTLgetvalue{\LACQUERSeqPrec}{effectivenss}{3}{\dtlcolumnindex{effectivenss}{Prec-S1}}
\DTLgetvalue{\LACQUERSeqNPVProp}{effectivenss}{6}{\dtlcolumnindex{effectivenss}{NPV-S1}}
\DTLgetvalue{\LACQUERRandRec}{effectivenss}{3}{\dtlcolumnindex{effectivenss}{Rec-R1}}
\DTLgetvalue{\LACQUERRandRecProp}{effectivenss}{6}{\dtlcolumnindex{effectivenss}{Rec-R1}}
\DTLgetvalue{\RipperNCBITrain}{effectivenss}{1}{\dtlcolumnindex{effectivenss}{Train}}
\DTLgetvalue{\ArmNCBITrain}{effectivenss}{2}{\dtlcolumnindex{effectivenss}{Train}}
\DTLgetvalue{\LACQUERNCBITrain}{effectivenss}{3}{\dtlcolumnindex{effectivenss}{Train}}
\DTLgetvalue{\LACQUERNCBITrain}{effectivenss}{3}{\dtlcolumnindex{effectivenss}{Train}}
\DTLgetvalue{\RIPPERPROPTrain}{effectivenss}{4}{\dtlcolumnindex{effectivenss}{Train}}
\DTLgetvalue{\ARMPROPTrain}{effectivenss}{5}{\dtlcolumnindex{effectivenss}{Train}}
\DTLgetvalue{\LACQUERPROPTrain}{effectivenss}{6}{\dtlcolumnindex{effectivenss}{Train}}
\DTLgetvalue{\LACQUERNCBIAvg}{effectivenss}{3}{\dtlcolumnindex{effectivenss}{P-Avg}}
\DTLgetvalue{\ARMNCBIAvg}{effectivenss}{2}{\dtlcolumnindex{effectivenss}{P-Avg}}
\DTLgetvalue{\LACQUERPROPAvg}{effectivenss}{6}{\dtlcolumnindex{effectivenss}{P-Avg}}
\newcommand{\LACQUERPredictMax}{839}

\def\UrlBreaks{\do\A\do\B\do\C\do\D\do\E\do\F\do\G\do\H\do\I\do\J
\do\K\do\L\do\M\do\N\do\O\do\P\do\Q\do\R\do\S\do\T\do\U\do\V
\do\W\do\X\do\Y\do\Z\do\[\do\\\do\]\do\^\do\_\do\`\do\a\do\b
\do\c\do\d\do\e\do\f\do\g\do\h\do\i\do\j\do\k\do\l\do\m\do\n
\do\o\do\p\do\q\do\r\do\s\do\t\do\u\do\v\do\w\do\x\do\y\do\z
\do\.\do\@\do\\\do\/\do\!\do\_\do\|\do\;\do\>\do\]\do\)\do\,
\do\?\do\'\do+\do\=\do\#} 
\maketitle

\section{Introduction}
\label{sec:introduction}

Software designers use data entry forms 
to collect inputs of users
who interact with software systems~\cite{jarrett2009forms, sears2003data}.
To correctly collect the necessary information from users,
designers typically define the completeness requirements of fields in data entry forms.
These completeness requirements specify the fields that are required or optional to fill for different types of users.

However, as the software system and the application requirements
change, data entry forms change too. Such changes may result in some
fields, previously marked as required, becoming inapplicable for
certain types of users. We call \emph{obsolete required fields} the fields whose ``required'' attribute does not remain valid with respect to the current application requirements. 
Although such fields are set as ``required'' in the form, the correct completeness requirement should be ``optional''.

When obsolete required fields are included in an input form, since the system usually has client-side validation checks~\cite{vassilakis2003framework} to ensure that
all the required fields have been filled in,
users are obliged to fill the required fields
with meaningless values (e.g., ``@'', ``n/a'') to be able to submit
the form~\cite{avidan2012record,kulyk2017sharing}.
We have observed this phenomenon both on a popular biomedical
information collection platform NCBI~\cite{barrett2012bioproject}, in
which more than half of the users have filled meaningless values in
required fields, and in a dataset provided by our industrial partner in
the financial domain.

Obsolete required fields represent an extra burden for the users,
costing additional time for filling the input form, and might lead to
users interrupting the data entry process,
with potential loss of business opportunities (e.g., a prospective
client giving up during the registration phase due to the complexity
of the input form).  Moreover, the meaningless values filled through
these obsolete required fields are then transferred to the software
system using them and may affect the overall data quality of the
system~\cite{mucslu2015preventing}.
For example, given a categorical field (which is an obsolete required field), 
the user can choose the first value in a combo box just to skip filling this field. 
Even though the value is chosen from the list of candidate values,
this value is meaningless since the field should not be filled at the beginning ~\cite{avidan2012record}. 
This value can be used as an input by machine learning-based tools (for example an automated form filling tool~\cite{belgacem2022machine}), which can then lead to more errors (e.g., wrongly predicting the values of some fields). 

To automatically relax completeness requirements and avoid meaningless
values, existing work has proposed adaptive form
tools~\cite{frank1998adaptive,bohoj2011adapforms,stromsted2018dynamic},
which enable form designers to set required fields as optional when
certain conditions hold.  These tools first require form designers to
define a \emph{complete and final} set of completeness requirements,
capturing the conditions for which a field should be required or
optional.  Then, they use intermediate representations such as
XML~\cite{bohoj2011adapforms} and dynamic condition response
graphs~\cite{stromsted2018dynamic} to represent the completeness
requirements rules and implement adaptive behaviors.
In addition, there are commercial tools (e.g., Gravity
	Forms~\cite{gravityforms}, Google Forms~\cite{googleforms}) that assist designers
	in designing adaptive forms,  where fields can be displayed or hidden based
	on the value of already filled fields in the form. 
	Similar to existing research approaches, these commercial tools assume that 
	designers already have a \emph{complete and
	final} set of completeness requirements describing the adaptive behaviour of the form during the design phase.

 However, due to
the complexity of the domain (with hundreds of fields) and the
evolution of the software, identifying a priori a comprehensive
set of completeness requirements is not a viable solution. Moreover,
even if they could be identified, such completeness requirements could
become quickly obsolete,  limiting the use of existing adaptive form tools.

To solve this problem, we propose \approach, a
\underline{L}earning-b\underline{A}sed \underline{C}ompleteness
re\underline{Q}\underline{U}ir\underline{E}ments
\underline{R}elaxation approach, to automatically learn the conditions
under which completeness requirements can be relaxed (i.e., when a
required field can become optional).  The basic idea of \approach is
to build machine learning models to learn the conditions under which
users had to fill meaningless values based on the data provided as
input in past data entry sessions (hereafter called \emph{historical
  input instances}).  Using these models, the already-filled fields in
a data entry form can then be used as features to predict whether a
required field should become optional for certain users.  \approach
can be used during the form filling process to refactor data entry forms by dynamically removing obsolete required
fields at run time, helping designers identify completeness requirements that should be relaxed.

\approach includes three phases: \emph{model building}, \emph{form
  filling relaxation}, and \emph{threshold determination}.  Given a
set of historical input instances, the model building phase identifies
the meaningless values filled by users and builds Bayesian
network (BN) models to represent the completeness requirement
dependencies among form  fields (i.e., the conditions upon which
users fill meaningless values).  To improve its learning ability,
\approach identifies also the cases where a required field is only
applicable for a small group of users; it uses the synthetic minority oversampling technique SMOTE to generate more instances on such fields for
effectively mining dependencies on them.
Once the
trained models are available, during the data entry session, the form
filling relaxation phase predicts the completeness requirement of a
target field based on the values of the already-filled fields and
their conditional dependencies in the trained models.  The predicted
completeness requirement of a field and the corresponding predicted
probability (endorsed based on a ``threshold'' automatically
determined) are then used to
implement adaptive behaviors of data entry forms.

The overall architecture of \approach has been inspired by  LAFF~\cite{belgacem2022machine}, our previous work on
	automated form filling of data entry forms.  The main similarities
	between these approaches derive from their shared challenges associated
	with the application domain (form filling). 
	These challenges include (1) the arbitrary filling order
	and (2) partially filled forms.  To address the
	first challenge, similar to  LAFF, we use BNs in order to mine
	the relationships between filled fields and the target field to
	avoid training a separate model for each filling order.  As for
	the second challenge, once again similar to LAFF, we use an endorser
	module to avoid providing inaccurate suggestions to the user when
	the form does not contain enough information for the
	model. More details about the similarities and differences between
	\approach and LAFF are provided in section~\ref{sec:related_work}.

We evaluated \approach using form filling records from both a public dataset and a proprietary dataset extracted from a production-grade enterprise information system in the financial domain.
The experimental results show that \approach can accurately relax the completeness requirements of required fields in data entry forms with a precision value between 0.76 and 0.90 when predicting the truly required fields.
In a sequential filling scenario, i.e., when users fill data entry forms in the default order determined by the form tab sequence,
\approach can prevent users from providing meaningless values in  20\% to
64\% of the cases, with a negative predictive value (representing the ability of \approach to correctly predict a field as ``optional'') between 0.72 and 0.91,
significantly outperforming state-of-the-art rule-based approaches by \SIrange{12}{70}{\pp} (with \si{\pp} = percentage
points) on the two datasets.
Furthermore, \approach is efficient; it takes at most
{\SI{\LACQUERPredictMax}{\milli\s}} to determine the completeness requirement of an input instance of the proprietary dataset.

To summarize, the main contributions of this paper are:
\begin{itemize}
	\item The \approach approach, which addresses the problem of
	automated completeness requirements relaxation --- an important challenge in designing data entry forms.
	To the
	best of our knowledge, 
	\approach is the first work to combine
	BNs with oversampling and a probability-based endorser to provide accurate completeness requirement suggestions.
\item An extensive evaluation assessing the effectiveness and
          efficiency of \approach and comparing it with
          state-of-the-art baselines\footnote{The implementation
              of \approach and the scripts used for evaluation are
              available at
              \url{https://figshare.com/s/0fdeae041e728e6d0a01}; see
              also section~\ref{sec:data-availability}.}.
\end{itemize}

The rest of the paper is organized as
follows. Section~\ref{sec:background} provides a motivating example
and explains the basic definitions of automated completeness
requirements relaxation and its
challenges. Section~\ref{sec:preliminaries} introduces the basic
machine learning algorithms used in this
paper. Section~\ref{sec:approach} describes the different steps and
the core algorithms of \approach. Section~\ref{sec:evaluation} reports
on the evaluation of \approach.  Section~\ref{sec:related_work}
surveys related work.  Section~\ref{sec:discussion} discusses the
usefulness and practical implication of \approach.
Section~\ref{sec:conclusion} concludes the paper.
 \section{Completeness Requirement Relaxation for Data Entry Forms}
\label{sec:background}

In this section, we introduce the concepts related to data entry
forms, provide a motivating example, precisely define the problem of automated
completeness requirement relaxation for data entry forms, and discuss its
challenges.

\subsection{Data Entry Forms}
\label{sec:data_entry_forms}

Data entry forms
are composed of fields of different types, such as
textual, numerical, and categorical.  Textual and numerical fields
collect free text and numerical values, respectively (e.g., the name
and the age of a private customer of an energy provider); categorical fields provide a list of
options from which users have to choose (e.g., nationality).  Form
developers can mark form fields either as \emph{required} or
\emph{optional}, depending on the importance of the information to be
collected.
This decision is made during the design phase of the form
based on the application completeness requirements.  Such requirements
capture the input data that shall be collected for certain types of
users; they are fulfilled by setting the required/optional property of
the corresponding fields in a data entry form.  In other words, the
required fields (also called mandatory
fields~\cite{seckler2014designing}) of a form collect input
information considered as important to the stakeholders who plan to
use the collected information; the absence of this information could
affect the application usage. On the contrary, optional fields collect information that is nice to have but whose absence is acceptable.
For example, an energy provider cannot open a customer account when the customer name
is missing; hence, the corresponding input field in a data entry form
should be marked as ``required''. At the same time, an energy provider does not
need to know the education level of a new private customer (though it could
be useful for profiling), so the corresponding input field can be marked
as ``optional''.

Some required fields can be further classified \emph{conditionally
required}, i.e., they are required only if certain conditions hold.
For example, the field ``marriage date'' is required only if the value
of the categorical field ``civil status'' is set to ``\texttt{married}''.  Data
entry forms that support ``conditionally required fields'' are generally called
\emph{adaptive forms}~\cite{bohoj2011adapforms} or context-sensitive
forms~\cite{avidan2012record}, since they exhibit adaptive behaviors
based on the values filled by users. More specifically, these types of
forms are programmed so that a field can be set from ``required'' to
``optional'' during the form-filling session, based on the input data;
a change of this property also toggles the visibility of the field
itself in the form.  Such adaptive behaviours make the data entry form
easier to use~\cite{avidan2012record}, since users can focus on the
right fields they need to fill in.

Before submitting a data entry form, the form usually conducts a
\emph{client-side validation} check~\cite{vassilakis2003framework} ---
using some scripting language or built-in features of the environment
where the form is visualized, like HTML attributes --- to ensure that
all the required fields have been filled in.

In this work, 
we consider a simple representation of an input form, 
with basic input fields that can have only a unique value that can be
selected or entered, such as a text box (e.g., \texttt{<input
	type="text">} or \texttt{<textarea>} in HTML), a drop-down menu (e.g, \texttt{<select>}
with single selection), or a radio
button (e.g., \texttt{<input type="radio"} in HTML).
This allows us to assume that a field can only have one
completeness requirement; in other words, a field cannot be
optional and required at the same time.

We do not support forms with more sophisticated controls or fields that can handle multiple selections 
(e.g., a checkbox group for multiple-choice answers or a drop-down
menu with multiple selection), 
as often found in surveys and questionnaires. 
Note that in this case a field can be both optional
and required at the same time, depending on the number of selected
values in the group\footnote{An example of complex
    input control is the case where users need to select at least
    three options from a multiple choice answer field (e.g., a
    checkbox group). Any option chosen before reaching the minimum
    number of selected values would be considered ``required'';
    however, the same option chosen after the first three would
    be considered ``optional''.
}. 
We plan to support this kind of 
complex controls as part of future work.

\subsection{Motivating Example}
\label{sec:motivating_example}
Data entry forms are difficult to
design~\cite{firmenich2012supporting} and subject to frequent
changes~\cite{yang2020managing}. These two aspects of data entry form
design and development negatively impact the way developers deal with
application completeness requirements in data entry forms. 

For example, let us consider a data entry form in an energy provider information
system, used for opening an account for business customers.  For
simplicity, we assume the form has only three required fields: ``Company type'' (categorical), ``Field of activity'' (categorical), and
``Tax ID'' (textual).  Sometime after the deployment of the initial
version of the system, the energy provider decides to support also the opening of
customer accounts for non-profit organizations (NPOs).  The developers update the
form by adding
\begin{inparaenum}[(a)]
  \item a new option ``\texttt{NPO}'' to the field
    ``Company type'', and
  \item additional fields denoting information required for
    NPOs.
\end{inparaenum}
After the deployment of the new form, a data entry operator of the
energy provider (i.e., the end-user
interfacing with the data entry form) notices a blocking situation
when filling in the form for an NPO. Specifically,
the form flags the
field ``Tax ID'' as required; however, the company
representative cannot provide one since the company is exempted from
paying taxes.  The clerk is then obliged to fill in the required field
with a meaningless value (e.g., ``@'') to pass the validation check
and be able to submit the form.  Several weeks later, after noticing
some issues in the generation of customers' reports, the data quality
division of the energy provider reviews the data collected through the
data entry form, detecting the presence of meaningless values.  A
subsequent meeting with IT analysts and developers reveals that those
values have been introduced because the data entry form design has not
been updated to take into account the new business requirements (i.e.,
opening accounts for NPOs) and the 
corresponding completeness requirements (i.e., some NPOs in certain fields of activity do not have a tax ID).  For
example, the current (but obsolete) form design always flags ``tax
ID'' as a \emph{required} field; however, when the ``Company type'' field
is set to ``\texttt{NPO}'' and the ``Field of activity''
field is either ``\texttt{charity}'' or ``\texttt{education}'', the field ``tax
ID'' should be \emph{optional}.

These meaningless values filled during form filling negatively
	affect data quality~\cite{avidan2012record}, 
	since they are considered as data entry errors and may lead to error propagation:
\begin{itemize}
	\item \textbf{Data entry errors}: 
	Users fill obsolete required fields with incorrect data (meaningless values) in order to proceed quickly in the workflow of the data-entry form~\cite{avidan2012record}. 
	\item \textbf{Error propagation}: Meaningless value errors can propagate and create more errors~\cite{mucslu2015preventing}, especially when these values are used in ML-based tools.
\end{itemize}
Meaningless value errors are difficult to identify because such values can pass all validation checks of the data entry form.
A business may establish the practice of using specific values (e.g., ``@'' and ``-1'') when users do not need to fill some fields, as in the aforementioned example. 
However, even in this case the data quality team needs to carefully check the filled fields to ensure that all the data entry operators follow this convention, which is a time-consuming process.

Currently, there are some simple but rather impractical solutions to
	address the issue of filling meaningless values, including
	rule-based solution and dictionary-based solution:

\begin{itemize}
\item \textbf{Rule-based solution}: 
This solution defines for each field some
rules capturing the conditions for which a required field can become
optional, based on the values of the other form fields.  
\item \textbf{Dictionary-based solution}: This solution sets all fields containing meaningless values as optional.
More specifically, the data quality division could first create a dictionary of meaningless values (e.g., ``@'', ``\$'').
Users can then use such values when a field is not applicable in a certain
form-filling scenario.
Finally, the data quality division could analyze the historical input
instances and mark 
a field as optional when users assign a value to it from the meaningless
values dictionary. Such information could then be used to refactor the data entry form, setting
the corresponding input field as optional.
\end{itemize}

However, the two solutions are not practical.
Given the evolving nature of software~\cite{wan2004relating,smr.1888},
the rule-based solution is not scalable and maintainable, especially when the
number of fields (and their possible values, for categorical fields)
increases.  Moreover, as is the case for our industrial partner, it
is difficult also for domain experts to formulate the completeness
requirement of new fields, since they have to decide the exact impact
of different field combinations on the new fields.
Regarding the dictionary-based solution,
the completeness requirement of a field usually depends on the values
of other filled fields~\cite{avidan2012record} (such as the
aforementioned example of Tax ID), and cannot be detected only by
looking at special/meaningless characters.
This simple solution cannot help domain experts identify these useful conditions.

Therefore, we contend it is necessary to develop automated methods to learn such
conditions directly from the data provided as input in past data
entry sessions, so that completeness requirements of form fields can
be automatically relaxed during new data entry sessions. Moreover, the
learned conditions could also help designers identify completeness
requirements that should be relaxed.

\subsection{Problem Definition}
\label{sec:definition}

In this paper, we deal with the problem of completeness requirement relaxation for data entry forms.
The problem can be informally defined as deciding whether a
required field in a form can be considered optional based on the
values of the other fields and the values provided as input in
previous data entry sessions for the same form.  We formally define this problem
as follows.

Let us assume we have a data entry form with $n$ fields $F=\{f_1, f_2, \dots, f_n\}$.
Taking into account the required/optional attribute of each field,
the set of fields can be partitioned into two groups: required fields
(denoted by $R$) and optional fields (denoted by $\bar{R}$),
where $ \bar{R} \cup R=F$ and $\bar{R} \cap R = \emptyset$.
Let $\mathit{VD}$ represent a value domain that excludes empty values.
Each field $f_i$ in $F$ can take a value from a domain
  $V_i$, where $V_i = \mathit{VD}_i$
if the field is required and   $V_i = \mathit{VD}_i \cup \bot$
if the field is optional ($\bot$ is a special element representing an empty value).

Let  $R^{c}\subseteq  R$ be the set of conditionally required fields,
which are required only when a certain condition $\mathit{Cond}$ is satisfied.
For a field $f_k\in R^{c}$, we define the condition $\mathit{Cond}_k$
as the conjunction of predicates over the value of some other fields; more
formally, $\mathit{Cond}_k=\bigwedge_{1 \leq i \leq n, i \neq k} h(f_i,
v_i^c )$, where $f_i \in F, v_i^c \in V_i$, and $h$ is a predicate
over the field $f_i$ with respect to the value $v_i^c$.

During form filling, at any time $t$ the fields can be partitioned into two groups:
fields that have been filled completely (denoted by $C_{t}$) and unfilled fields (denoted by $\bar{C_{t}}$); 
let $G$ be the operation that extracts a field from a form during form filling $G(F)= f$, such that  $(f \in C_t)\lor (f\in \bar{C_{t}})$
 and $C_{t} \cap \bar{C_{t}}= \emptyset$.
By taking into account also the required/optional attribute,
we have: filled required fields $(C_{t}\cap R)$, filled optional fields $(C_{t}\cap \bar{R})$, 
unfilled required fields $(\bar{C_{t}} \cap R)$, and unfilled optional fields $(\bar{C_{t}} \cap \bar{R})$.

When a form is about to be submitted (e.g., to be stored in
a database), 
we define an \emph{input instance} of the form to be
$I^F=\{\langle f_1, v_1 \rangle, \dots, \langle f_n, v_n\rangle\}$ with
$f_i \in F$ and $v_i \in V_i$;
we use the subscript $t_j$ as in $I^F_{t_j}$ to denote that 
the input instance $I^F$ was submitted at time $t_j$.
We use the notation $I^F(t)$ to represent the set of \emph{historical input instances} of the form
that have been submitted up to a certain time
instant $t$;
$I^F(t)=\{I^F_{t_{i}}, I^F_{t_{j}}, \dots, I^F_{t_{k}}\}$, 
where $t_i < t_j < t_k < t$. 
Hereafter, we drop the superscript $F$ when it is clear from the context.

The completeness requirement relaxation problem can be defined as follows.
Given a partially filled form $F=\{f_1, f_2, \dots, f_n\}$
  for which, at time $t$, we know $\bar{C_{t}} \neq
  \emptyset$, $C_{t}$, and $R^c$, 
a set of historical input instances $I^F(t)$, 
and a target field $f_{p}\in (R^{c}\cap \bar{C_{t}})$ to fill,
with $p \in 1\dots n$,
we want to build a model $M$ predicting whether, at time $t$, $f_p$ should become optional based on $C_{t}$ and $I^F(t)$. 

\begin{figure}[tb]
	\centering{
\tikzstyle{field} = [rectangle, minimum width=1.2cm, minimum height=0.7cm, text centered, text width=2.3cm, draw=black, fill= white!30]

\tikzstyle{label} = [rectangle, minimum width=1.7cm, minimum height=0.35cm, text centered, text width=2.5cm, align=right, draw=white, fill= white!30]

\tikzstyle{title_node} = [rectangle, minimum width=0.5cm, minimum height=0.35cm, text centered, text width=4.5cm, draw=white, fill= white!30]

\tikzstyle{text_node} = [rectangle, minimum width=2.5cm, minimum height=0.7cm, text centered, text width=2.5cm, fill=white!30]

\tikzstyle{round_rect} = [rectangle, rounded corners, minimum width=1.5cm, minimum height=0.5 cm,text centered, draw=black, fill=white!30]

\newcommand{\DrawTriangle}[1][]{\begin{tikzpicture}[overlay,remember picture]
	\filldraw[fill=black] (90:1.2ex) -- (170:0.7ex) -- (10:0.7ex) --cycle;
	\end{tikzpicture}
}

\resizebox{1\textwidth}{!}{
\begin{tikzpicture}[node distance=3mm, >=latex]
	
	\node (tb1) [field]{Wish};
	\node (f1) [label, left= of tb1] {Company Name};
	\node (tb2) [field, below=of tb1, text width=1.1cm, minimum width=1.1cm, xshift=-0.6cm] {20};
	\node (f2) [label, left= of tb2] {Monthly revenue};
	\node (f2_unit) [label, right= of tb2, xshift=-2mm, align=left, text width=1.3cm] {\emph{k euro}};
	\node (tb3) [field, below=of tb2,xshift=0.6cm] {NPO};
	\node (f3) [label, left= of tb3] {Company type};
	\node (tb4) [field, below=of tb3] {education};
	\node (f4) [label, left= of tb4] {Field of activity};
	\node (tb5) [field, below=of tb4, text=red,draw= red] {};
	\node (f5) [label, left= of tb5, text=red] {Tax ID};
	
	\node (icon_tb3) [isosceles triangle,
	isosceles triangle apex angle=60, draw, rotate=270, fill=gray!120, minimum size =0.1cm, right=of tb3, xshift=-0.25cm, yshift=-0.6cm]{};
	
	\node (icon_tb4) [isosceles triangle,
	isosceles triangle apex angle=60, draw, rotate=270, fill=gray!120, minimum size =0.1cm, right=of tb4, xshift=-0.25cm, yshift=-0.6cm]{};
		
	\node (submit) [round_rect, below= of tb5, xshift=0.35cm, minimum height=0.7cm, text centered, text width=1.6cm, fill=gray!30] {Submit};
	
	\node (cancel) [round_rect, left= of submit, xshift=-0.35cm, minimum height=0.7cm, text centered, text width=1.6cm, fill=gray!30] {Cancel};
	
	\node (ui) [draw=none, fit= (tb1) (tb2) (tb3)(tb4)(tb5)(f1)(f2)(f3)(f4)(f5)(submit) ] {};
	
	\node (title)[title_node, font=\fontsize{12}{0}\selectfont, above=of ui] {\textbf{Data entry form \emph{F}}};
	
	\node (form) [draw=black, fit= (ui)(title)] {};
		
	\draw [decorate, decoration={brace, aspect=0.37, amplitude=10pt, raise=4pt}, yshift=0pt] (tb1.north east) -- (tb4.south east)  node [black, xshift=0cm] { };
	
	\node (model) [round_rect, right= of tb3, xshift=2cm, minimum height=1cm, text centered, text width=2.3cm, yshift=1cm] {\textbf{Model $\mathbf{M}$}};
	
	\draw [arrow,->] (tb3) ++(1.8,1) -- (model) node[midway, text width=2.5cm, text centered] {$C_t$\\ filled fields} ;
	
	\draw [arrow,->] (model) |- (tb5) node[anchor=north east, above, pos=0.80] {predict};

	\node (tab1) [shape=rectangle, draw=white, right= of submit, xshift=3.9cm, yshift=1cm] {
		\scriptsize
		
		\begin{tabular}{cccccc}
			\toprule
			
			\textbf{$f_1$: Company}&\textbf{$f_2$: Monthly}&\multirow{2}{*}{\textbf{$f_3$: Company type}}&\textbf{$f_4$: Field of }&\multirow{2}{*}{\textbf{$f_5$: Tax ID}}&\multirow{3}{*}{\textbf{Submission}}\\
			\textbf{Name}&\textbf{revenue}&&\textbf{ activity}&&\\
			\emph{(Textual)}&\emph{(Numerical)}&\emph{(Categorical)}&\emph{(Categorical)}&\emph{(Textual)}\\
			\midrule
			
			UCI &20&Large enterprise&Real estate&T190&20180101194321 \\
			\midrule
			KDL&21& Large enterprise&Manufacturing&T201 &20180101194723 \\ 
			\midrule
			...&...&  ...&...& ...  \\ 
			\midrule
			UNI &39&NPO &Education&n/a&20180102132016 \\

			\bottomrule
		\end{tabular}
		
	};
	
	\draw [arrow,->] (submit)  -- ++(5.1, 0)(tab1) node[above, pos=0.4, text width=2.5cm, text centered] {submission} ;
	
	\node(history)[text_node, above= of tab1, text width=5cm, yshift=-0.4cm]{Historical input instances $I^F(t)$};
	
	\draw [arrow,->] (history)  |- (model) node[above, pos=0.75, text width=2.5cm, text centered] {train} ;

\end{tikzpicture}
}
 }
	\caption{The Automated Form Filling  Relaxation Problem
	}
	\label{fig:def}
      \end{figure}
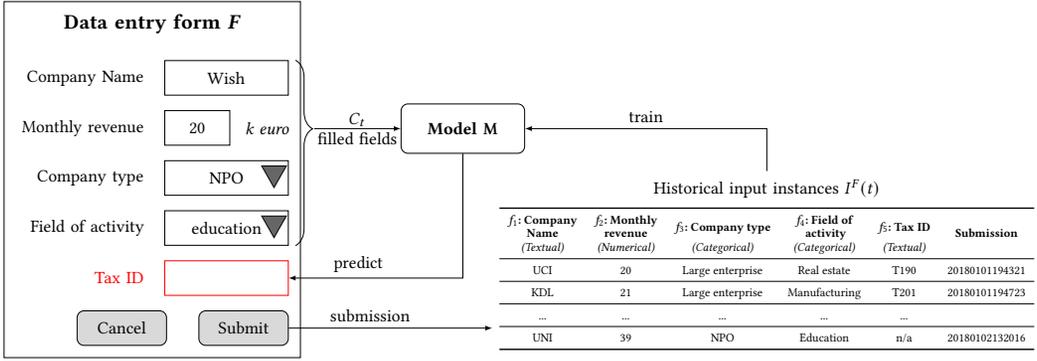

\subsubsection*{Framing the problem definition scope}

In this problem definition, our goal is \emph{to relax the completeness
  requirements of a form} by determining which obsolete required
  fields should become optional to avoid filling meaningless values.
  We do not consider the case in which optional fields could become
  required; we leave extending \approach to automatically decide the
  completeness requirement of all fields as part of future work.

Moreover, as mentioned in the motivating example, 
in this definition, we mainly focus on the case of filling data entry
forms from scratch.
We do not consider the case in which an existing instance in the
database is updated (including an update of the timestamp); 
for example, following our motivating example, if a company changes its ``Field of activity'' to ``\texttt{charity}'',
then some fields like ``tax ID'' may become optional and do not need to be filled.
\approach can be adapted to support this scenario and check if the completeness requirement 
of some fields need to be changed; we also leave this adaption as part of future work.

\subsubsection*{Application to the running example}
\figurename~\ref{fig:def} is an example of a data entry form used 
to fill information needed to open an account for business customers
with an energy provider
The form $F$ is composed of five fields, including $f_1$:``Company
Name'', $f_2$:``Monthly revenue'',
$f_3$:``Company type'', $f_4$:``Field of activity'', and $f_5$:``Tax ID''.
All the fields are initially required (i.e., $R=\{f_1,f_2,f_3,f_4, f_5\}$).
Values filled in these fields are then stored in a database.
An example of the database is shown on the right side of  \figurename~\ref{fig:def}.
These values are collected during the data entry session with an automatically recorded timestamp indicating the submission 
time. 
Each row in the database represents an input instance (e.g.,
$I^F_{20180101194321}=\{\langle \textit{``Company Name''}, \textit{UCI} \rangle, \dots, \langle \textit{``Tax ID''}, \textit{T190}\rangle\}$), where the column name corresponds to the field name in the form.
The mapping can be obtained from the existing software design documentation or software implementation~\cite{belgacem2022machine}.
Using the data collected from different users, we can build a model $M$ to learn 
possible relationships of completeness requirement between different fields.
Let us assume a scenario where during the creation of a customer account using $F$,
the energy provider clerk has entered \emph{Wish}, \emph{20}, 
\emph{NPO}, and \emph{education} for fields $f_1$ to $f_4$, respectively.
The field $f_5$ (``Tax ID'') is the next field to be filled. 
Our goal is to automatically decide if field $f_5$ is required or not 
based on the values filled in fields $f_1$ to $f_4$.

\subsection{Towards adaptive forms: challenges}
\label{sec:challenges}

Several tools  for adaptive forms have been
 proposed~\cite{frank1998adaptive,bohoj2011adapforms,stromsted2018dynamic}.
These approaches use intermediate representations such as XML~\cite{bohoj2011adapforms}
 and dynamic condition response graphs ~\cite{stromsted2018dynamic} 
 to represent the completeness requirements rules and implement adaptive behaviours.
 Existing tools for adaptive forms
 usually assume that form designers already have, during the design
 phase, a \emph{complete and final} set of
 completeness requirements, capturing the conditions for which a field
 should be required or optional.

 However, this assumption is not valid in real-world applications.  On
 one hand, data entry forms are not easy to
 design~\cite{firmenich2012supporting}.  Data entry forms need to
 reflect the data that need to be filled in an application domain.
 Due to time pressure and the complexity of the domain (e.g., the
 number of fields needed to be filled and their interrelation), it is
 difficult to identify all the completeness requirements when
 designing the data entry
 form~\cite{dalpiaz2019detecting,aldekhail2017intelligent}.
 On the other hand, data entry forms
 are subject to change: a recent study~\cite{yang2020managing} has shown that  49\% of web applications will modify their data
 constraints in a future version.  The frequent changes in data
 constraints may also make the existing completeness requirements
 obsolete.

Hence the main challenge is how to create adaptive forms 
when the set of completeness requirements representing the adaptive
behaviour of a form is incomplete and evolving.

 \section{Preliminaries}
\label{sec:preliminaries}

Before illustrating our approach, we first briefly introduce two basic
machine-learning algorithms we rely on.

\subsection{Bayesian Networks}
\label{sec:bayesian-networks}

Bayesian networks (BNs) are probabilistic graphical models (PGM) in
which a set of random variables and their conditional dependencies are
encoded as  a directed acyclic graph: nodes correspond to random
variables and edges correspond to conditional probabilities.

The use of BNs for supervised learning~\cite{friedman1997bayesian}
typically consists of two phases: structure learning and variable inference.

During \emph{structure learning}, the graphical structure of the BN is
automatically learned from a training set.  First, the conditional
probability between any two random variables is computed.  
Based on
these probabilities, optimization-based search (e.g., hill
climbing~\cite{gamez2011learning}) is applied to search the graphical structure.
The search algorithm initializes a random structure,
and then iteratively adds or deletes its nodes and edges to generate new structures.
For each new structure, the search algorithm calculates a fitness
function (e.g., Bayesian information criterion,
BIC~\cite{raftery1995bayesian}) based on the nodes' conditional
probabilities and on Bayes' theorem~\cite{friedman1997bayesian}.
Structure learning stops when it finds a graphical structure that minimizes the fitness function.

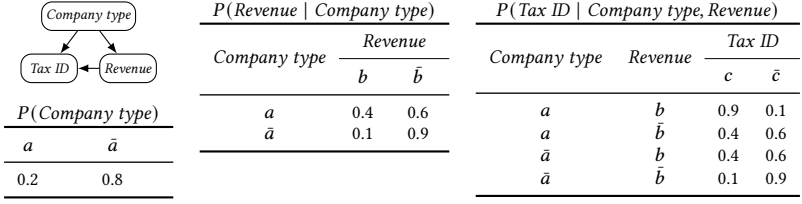
\begin{figure}[tb]
	
\tikzstyle{round_rect} = [rectangle, rounded corners, minimum width=1cm, minimum height=0.5 cm,text centered, draw=black, fill=white!30]
\tikzstyle{rect} = [rectangle, minimum width=1.5cm, minimum height=0.5cm, text centered, text width=1.5cm, draw=black, fill=white!30]
\tikzstyle{text_node} = [rectangle, minimum width=0.5cm, minimum height=0.35cm, text centered, text width=4cm, draw=white, fill= white!30]

\begin{minipage}[t]{\linewidth}
  \centering
  \scriptsize
\begin{tabular}{ccc}
\begin{tikzpicture}[baseline,scale=0.7, every node/.style={scale=0.75},node distance=3mm, >=latex]

\node (A) [round_rect] {$\mathit{Company \ type}$};
\node (B) [round_rect, below= of A,xshift= 0.7cm] {$\mathit{Revenue}$};
\node (C) [round_rect, below= of A,xshift=-0.7cm] {$\mathit{Tax \ ID}$};

\draw [arrow,->] (A) -- (C);
\draw [arrow,->] (A) -- (B);
\draw [arrow,->] (B) -- (C);

\end{tikzpicture}

&\multirow{3}{*}{\begin{tabular}[b]{ c c c}
          \multicolumn{3}{c}{$P(\mathit{Revenue} \mid \mathit{Company \ type})$}\\
          \toprule
        \multirow{2.5}{*}{$\mathit{Company \ type}$} & \multicolumn{2}{c}{$\mathit{Revenue}$} \\ 
	\cmidrule{2-3}
	&$b$&$\Bar{b}$ \\
	\midrule
	$a$  & 0.4 & 0.6 \\
	$\Bar{a}$ & 0.1 & 0.9 \\
	\bottomrule
	\end{tabular}}
    &\multirow{3}{*}{\begin{tabular}[b]{c  c c  c}
          \multicolumn{4}{c}{$P(\mathit{Tax \ ID}\mid \mathit{Company \ type}, \mathit{Revenue})$} \\
          \toprule
	\multirow{2.5}{*}{$\mathit{Company \ type}$} & \multirow{2.5}{*}{$\mathit{Revenue}$} & \multicolumn{2}{c}{$\mathit{Tax \ ID}$} \\             
	\cmidrule{3-4}
	&&$c$&$\Bar{c}$\\
	\midrule
	$a $& $b $& 0.9 & 0.1  \\
	$a $& $\Bar{b}$ & 0.4 & 0.6 \\
	$\bar{a}$&$b$& 0.4 & 0.6 \\
	$\Bar{a}$ &$\Bar{b}$ & 0.1 & 0.9 \\
	\bottomrule
	\end{tabular}}\\ &&\\ 
   \begin{tabular}[b]{ c c}
    \multicolumn{2}{c}{$P(\mathit{Company \ type})$}\\
	\toprule
	$a$&$\Bar{a}$ \\
	\midrule
	0.2 & 0.8 \\
	\bottomrule
   \end{tabular}
&&\\    
\end{tabular}
\end{minipage}

 	\caption{An Example of BN and the Probability
		Functions of its Nodes}
	\label{fig:bayesian_network}
\end{figure}

\figurename~\ref{fig:bayesian_network} shows an
example of the BN structure learned based on the data submitted by the data entry form used in our example in section~\ref{sec:motivating_example}.
This BN contains three nodes corresponding to three fields in the data entry form: variable $\mathit{Revenue}$
depends on variable $\mathit{Company \ type}$; variable $\mathit{Tax \ ID}$ depends on variables $\mathit{Company \ type}$ and
$Revenue$. 
For simplicity, we assume that the three variables are Boolean where $a$, $b$, and $c$ denote 
that fields $\mathit{Company \ type}$, $\mathit{Revenue}$, and $\mathit{Tax \ ID}$  are ``required'' respectively, and
$\bar{a}$, $\bar{b}$, and $\bar{c}$ denote that these fields are optional. 

In the PGM, each node is associated with a probability function
(in this case, encoded as a table), which represents the conditional
probability between the node and its parent(s).  For example, in
\figurename~\ref{fig:bayesian_network} each variable has two values;
the probability table for $\mathit{Revenue}$ reflects the conditional probability
$P(\mathit{Revenue}\mid \mathit{Company \ type})$ between $\mathit{Company \ type}$ and $\mathit{Revenue}$ on these values.

\emph{Variable inference} infers unobserved variables from the
observed variables and the graphical structure of the BN using Bayes' theorem~\cite{friedman1997bayesian}.
For example, we can infer the probability of $\mathit{Tax \ ID}$ to be required  (i.e., $\mathit{Tax \ ID} =c$) when the completeness requirement of $\mathit{Company \ type}$ is  required  (denoted by $P(c \mid a)$) as follows:
\begin{equation*}
	\begin{aligned}
		P(c \mid a) &= \frac{P(a, c)}{P(a)} = \frac{P(a, b, c) + P(a, \overline{b}, c)}{P(a)} \\
		&= \frac{P(c \mid a, b) P(b \mid a)  P(a) + P(c \mid a, \overline{b}) P(\overline{b} \mid a)  P(a)}{P(a)} \\
		&= \frac{0.9*0.4*0.2 + 0.4*0.6*0.2}{0.2} =0.6
	\end{aligned}
\end{equation*}

BNs have been initially proposed for learning dependencies among discrete
random variables.  They are also robust when dealing with missing
observed variables; more specifically, variable inference can be
conducted when some conditionally independent observed variables are
missing~\cite{friedman1997bayesian}.
Recently, they have been applied in the context of automated form filling~\cite{belgacem2022machine}.

\subsection{Synthetic Minority Oversampling Technique (SMOTE)}
\label{sec:smote}
A frequently encountered problem for training machine learning models using real-world data 
is that the number of instances per class can be imbalanced~\cite{song2018comprehensive,malhotra2017empirical}. 
To address this problem,
many imbalanced learning approaches have been proposed in the
literature.
One of them is SMOTE~\cite{chawla2002smote}; it uses an oversampling method to modify the class distribution in a dataset 
(i.e., the ratio between instances in different classes). 
It synthesizes new minority class instances to improve the learning ability of machine learning algorithms on the minority class. 
SMOTE conducts the instance synthesis by means of interpolation between near neighbors.
Initially, each instance in the dataset is represented as a feature vector.
SMOTE starts by randomly selecting a minority class instance $i$ from the dataset. 
It determines the $k$ nearest neighbors of $i$ from the remaining instances in the minority class by calculating their distance (e.g., the Euler distance) based on their feature vectors. 
SMOTE synthesizes new instances using $n$ instances randomly selected from the $k$ neighbors.
The selection is random to increase the diversity of the generated new instances.
For each selected instance,
SMOTE computes a ``difference vector'' that represents the difference
of the feature vectors between the selected instance and instance $i$.
SMOTE synthesizes new instances by adding an offset to the feature
vector of instance $i$,
where the offset is the product of the difference vector with a random number between 0 and 1.
SMOTE stops generating new instances until a predefined condition is satisfied 
(e.g., the ratio of instances in the majority and minority classes is the same).

\begin{figure}[tb]

\newcommand{\tikzmark}[2]{
	\tikz[overlay,remember picture,baseline] 
	\node [anchor=base] (#1) {$#2$};}

\newcommand{\DrawVLine}[3][]{\begin{tikzpicture}[overlay,remember picture]
		\draw[shorten <=0.3ex, #1] (#2.north) -- (#3.south);
	\end{tikzpicture}
}

\newcommand{\DrawCircle}[1][]{\begin{tikzpicture}[overlay,remember picture]
		\draw[black,fill=black] (0,0.5ex) circle (0.5ex);
	\end{tikzpicture}
}

\newcommand{\DrawRec}[1][]{\begin{tikzpicture}[overlay,remember picture]
		\draw[black,fill=black] (-0.5ex,0) rectangle ++(1ex, 1ex);
	\end{tikzpicture}
}

\newcommand{\DrawTriangle}[1][]{\begin{tikzpicture}[overlay,remember picture]
		\filldraw[fill=black] (90:1.2ex) -- (170:0.7ex) -- (10:0.7ex) --cycle;
	\end{tikzpicture}
}

\tikzstyle{round_rect} = [rectangle, rounded corners, minimum width=1.5cm, minimum height=0.5 cm,text centered, draw=black, fill=white!30]

\tikzstyle{rect} = [rectangle, minimum width=1.5cm, minimum height=0.8cm, text centered, text width=1.5cm, draw=black, fill=white!30]

\tikzstyle{text_node} = [rectangle, minimum width=0.5cm, minimum height=0.35cm, text centered, text width=4cm, draw=white, fill= white!30]
\resizebox{0.68\textwidth}{!}{
	\begin{tikzpicture}
		
		\node (tab1) [shape=rectangle,draw=white] {
			\scriptsize
\begin{tabular}{ccc}
			\toprule
			
			\textbf{}&\textbf{$f_2$: Monthly}&\multirow{2}{*}{\textbf{$f_3$: Target}}\\
			\textbf{}&\textbf{revenue}&{}\\
			
			\midrule
			
			i1&39&Optional\\
			i2&42&Optional\\
			i3&25&Optional\\
			i4&100&Required\\
			i5&150&Required\\
			i6&200&Required\\
			i7&400&Required\\
				\bottomrule
			\end{tabular}
};

	\node (tab2) [shape=rectangle,draw=white, right =of  tab1] {
		\scriptsize

\begin{tabular}{ccc}
	\toprule
	
	\textbf{}&\textbf{$f_2$: Monthly}&\multirow{2}{*}{\textbf{$f_3$: Target}}\\
	\textbf{}&\textbf{revenue}&{}\\
	
	\midrule
	
	i1&39&Optional\\
	i2&42&Optional\\
	i3&25&Optional\\
	i4&100&Required\\
	i5&150&Required\\
	i6&200&Required\\
	i7&400&Required\\
    i8&40&Optional\\

			\bottomrule
		\end{tabular}

	};
\draw[arrow,->] (tab1)--node[anchor=south] {\scriptsize SMOTE}(tab2);
\end{tikzpicture}}
 	\caption{An Example of SMOTE Interpolation}
	\label{fig:smote-pre}
\end{figure}

\figurename~\ref{fig:smote-pre} illustrates the application of SMOTE to create new minority class instances.
As shown in the table on the right, instances $i_1$,$i_2$, and $i_3$ belong to the minority class ``Optional'' of our target field.
As a preliminary step, SMOTE computes the Euclidean distance between all the minority instances:  $d({i_1},{i_2})= \sqrt {\left( 39-42\right)^2} = 3$, $d({i_1},{i_3})= \sqrt {\left( 39-25\right)^2} = 14$, and $d({i_2},{i_3})= \sqrt {\left( 42-25\right)^2} = 17$.
SMOTE starts by randomly picking one instance from the minority class
(e.g., $i_2$). 
Assuming that the value of $k$ is equal to 1, 
SMOTE selects the nearest instance to $i_2$, which in our example is the instance $i_1$. 
In order to create a new instance $i_8$, SMOTE computes the $\mathit{Difference \  vector}$ based on the feature vectors $\mathit{Monthly \ revenue}_{i_2} $ and $\mathit{Monthly \ revenue}_{i_1}$, and multiplies it by a random value $\lambda$ between 0 and 1.
The value of the ``Monthly revenue'' column in the synthetically created instance  $i_8$ is equal to 
$\mathit{Monthly \ revenue}_{i_2}$+ $\mathit{Difference \ vector}$.
In our example, assuming that the value of $\lambda$ is equal to 0.7, the new value of the ``Monthly revenue'' field for $i_8$ is equal to 
$42+((39 - 42) * 0.7)= 40$.

 \section{Approach}
\label{sec:approach}

\begin{figure}[tb]
	\centering
	
\tikzstyle{rounded_rect} = [rectangle, rounded corners, minimum width=2.5cm, minimum height=1cm,text centered, draw=black, fill=white!30]

\tikzstyle{rect} = [rectangle, minimum width=2.5cm, minimum height=1cm,  text width=2.5cm, draw=black,align=center, fill=white!30]

\tikzstyle{text_node} = [rectangle, minimum width=0.5cm, minimum height=0.5mm, text centered, text width=1cm, draw=white, fill= white!30]
  
\tikzstyle{list_box} = [rect,align=left ,minimum width=3.2cm,text width=3.2cm, minimum height= 0.5cm]

\resizebox{0.56\textwidth}{!}{
\begin{tikzpicture}[node distance=7mm, >=latex]
  \node (nd1) [rect,align=center] {Historical input instances};
  \node (nd2) [rounded_rect, below= of nd1, yshift=-0.5cm] {Pre-processing};
  \node (nd3) [rounded_rect, below= of nd2] {Model building};
  \node (big_box1)[draw=black, fit={(nd2)(nd3)},minimum width=2.9cm,minimum height=3.2cm,line width=0.4mm]{};
  
  \node (nd4) [rounded_rect, right= of nd2,xshift= 0.8cm] {Pre-processing};
  \node (nd5) [rounded_rect, below= of nd4] {Prediction};
  \node (nd6) [rounded_rect, right= of nd5, xshift=1.45cm] {Endorsing}; 
  \node (big_box2) [draw=black, line width=0.4mm,fit={(nd4)(nd5)(nd6)},minimum width=7.7cm, minimum height=3.2cm]{};
  
  \node (user)[inner sep=0pt, above= of nd4,xshift=-0.9cm,yshift=-0.3cm]
  {\includegraphics[width=.03\textwidth]{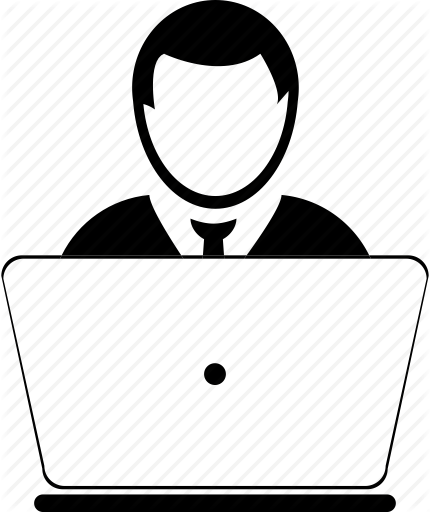}};
  \node (current_input) [rect] at (nd4 |- nd1) {Current input};
  
  \node (list_box_p1)[list_box, text width=2.8cm, minimum width=2.8cm, minimum height= 1cm] at (nd6 |- current_input) {Rquired   \hfill 0.70\\ Optional \hfill \ \ \ 0.30};

  \node (tuning) [rounded_rect, below= of nd3, xshift=4cm] {Threshold determination}; 
  \node (big_box3)[draw=black, fit={(tuning)},minimum width=11.7cm,minimum height=1.5cm,line width=0.4mm, xshift=10.5,yshift=-1]{};  
  \draw[arrow,->] (nd1)--(nd2);
  \draw[arrow,->] (nd2)--(nd3);
  \draw[arrow,->] (nd3)--node [anchor=south] {BNs}(nd5);
  \draw[arrow,->] (nd4)--(nd5);
  \draw[arrow,->] (nd5) --node [anchor=south,text width = 2cm,align=center] {Probabilistic distribution}(nd6);
  \draw [arrow,->] (current_input) -- (nd4);
 \draw [arrow,->] (nd6)-- node [anchor=east,text width = 3cm,xshift=1.2cm]{Suggestions}(list_box_p1);
\draw[arrow,->](nd3)  ++(0,-0.5)--++(0,-0.5)(big_box3)[anchor=south];
  \draw[arrow,->](big_box3)++(4.3,0.8)--++(0,0.45) (nd6);

  \node[shape=circle, draw, fill=gray, opacity=.2, text opacity=1, inner sep=0.5pt, left=of big_box1, xshift=0.6cm, yshift=1.3cm] (charA) {A};
  \node[shape=circle, draw, fill=gray, opacity=.2, text opacity=1, inner sep=0.5pt, left=of big_box2, xshift=0.6cm, yshift=1.3cm] (charB) {B};
  \node[shape=circle, draw, fill=gray, opacity=.2, text opacity=1, inner sep=0.5pt, left=of big_box3, xshift=0.6cm, yshift=0.5cm] (charC) {C};

  \node(MB_phase) [text_node, below=of big_box3, yshift=0.5cm, text width=11.7cm, align=left]{\myblock{A} Model Building Phase, \myblock{B} Form Filling Relaxation Phase, and \myblock{C}~Threshold Determination Phase};

\end{tikzpicture}
}

 	\caption{Main Steps of the \approach Approach}
	\label{fig:lacquer-framework}
\end{figure}

In this section, we present our machine learning approach for data entry form relaxation named \approach(\underline{L}earning-b\underline{A}sed \underline{C}ompleteness re\underline{Q}\underline{U}ir\underline{E}ments \underline{R}elaxation).

As shown in \figurename~\ref{fig:lacquer-framework},
\approach includes three phases: \emph{model building}, \emph{form filling relaxation}, and \emph{threshold determination}. 
\approach preprocesses the historical input instances related to a data entry form and identifies the \emph{meaningless} values in them. 
The historical input instances are divided in two parts: historical input instances for training (training input instances) and historical input instances for tuning (tuning input instances) used for threshold determination.
In the first phase, \approach builds BN models on the preprocessed training input instances
to represent the completeness requirement dependencies between form  fields. 
This phase occurs offline before deploying \approach as a completeness requirement relaxation tool for data entry.
The form filling relaxation phase occurs during the data entry session and assumes that all the models have been built.
During this phase,
given a target field, \approach selects the BN model related to the target 
from all the BN models
and predicts the completeness requirement of the target, 
 taking into account the values of the filled fields captured during the form filling process.
To improve prediction accuracy, \approach includes an endorser module 
that seeks to only provide users with predictions 
whose confidence level is higher than a minimum threshold.
The value of the threshold is automatically determined in the threshold determination phase.

\approach is inspired by our previous work on automated form filling~\cite{belgacem2022machine};
the main differences between the two approaches are discussed in section~\ref{sec:related_work}.

\subsection{Pre-processing}
\label{sec:pre-processing}

The first two phases of \approach include a preprocessing step 
to improve the quality of the data in historical input instances as
well as the current input instance.
As mentioned in section~\ref{sec:data_entry_forms}, data entry forms
can contain fields that are not applicable to certain users;
this is the main cause of the presence of \emph{missing values} and \emph{meaningless values} in historical input instances.
\emph{Missing values} occur when 
users skip filling an (optional) field during form filling.
A \emph{meaningless value} is defined as any value filled into a form field 
that can be accepted during the validation check 
but does not conform with the semantics of the field. 
For example, given a data entry form with a textual field ``Tax ID'', 
if a user fills ``n/a'' in this field,
the value can be accepted during the submission of the
instance\footnote{We assume the validation check does not check for
  the well-formedness of the string corresponding to the Tax ID.};
however, it should be deemed meaningless 
since ``n/a'' does not represent an actual ``Tax ID''.

For missing values, we replace them with a dummy value ``Optional'' in the corresponding field.
As for the meaningless values, 
we first create a dictionary containing possible meaningless values based on domain knowledge. 
This dictionary is used to match possible meaningless values in historical input instances;
we replace the matched values with ``Optional''. 
The rationale for this strategy is that it is common practice, within
an enterprise, to suggest data entry operators some specific keywords 
when a field is not applicable for them. 
For example, our industrial partner recommends users to fill such fields 
with special characters such as ``@'' and ``\$''.
The overarching intuition behind replacing missing values and
meaningless values with ``Optional'' is that, when data entry operators
skip filling a field (resulting in a missing value in the form) or put a meaningless value,
it usually means that this field is not applicable in the current context.

After detecting missing values and meaningless values,
we preprocess other filled values.
For textual fields, we replace all valid values with a dummy value ``Required'',
reflecting the fact that data entry operators deemed these fields to be 
applicable.
After preprocessing, all values in textual fields are therefore either ``Required'' and ``Optional''
to help the model learn the completeness requirement based on this abstract presentation. 
Numerical fields can be important to decide the completeness requirement of other fields. 
For example, companies reaching a certain monthly revenue can have some specific required fields. 
For this reason, we apply data discretization to numerical fields to
reduce the number of unique numeric values. 
Each numeric value is represented as an interval,
which is determined using the widely used discretization method based on information gain analysis ~\cite{breiman1984classification}.
We do not preprocess categorical fields 
since they have a finite number of candidate values.
We keep the original values of categorical fields 
since users who select the same category value may share common required information. 
At last, we delete all the fields that are consistently marked as ``Required'' or ``Optional'',
because such fields do not provide any discriminative knowledge to the model.

During the data entry session, similar preprocessing steps are applied.
We skip values filled in fields that were removed in historical input instances.
We replace values in textual fields with ``Required'' and
``Optional'', as described above.
We also map numerical values onto intervals and keep values in categorical fields. 

The historical input instances are then divided in two parts that will be used separately for training (training input instances)
and for the threshold determination (tuning input instances).

\subsubsection*{Application to the running example}
\begin{figure}[tbp!]
	\centering

\newcommand{\tikzmark}[2]{
	\tikz[overlay,remember picture,baseline] 
	\node [anchor=base] (#1) {$#2$};}

\newcommand{\DrawVLine}[3][]{\begin{tikzpicture}[overlay,remember picture]
		\draw[shorten <=0.3ex, #1] (#2.north) -- (#3.south);
	\end{tikzpicture}
}

\newcommand{\DrawCircle}[1][]{\begin{tikzpicture}[overlay,remember picture]
		\draw[black,fill=black] (0,0.5ex) circle (0.5ex);
	\end{tikzpicture}
}

\newcommand{\DrawRec}[1][]{\begin{tikzpicture}[overlay,remember picture]
		\draw[black,fill=black] (-0.5ex,0) rectangle ++(1ex, 1ex);
	\end{tikzpicture}
}

\newcommand{\DrawTriangle}[1][]{\begin{tikzpicture}[overlay,remember picture]
		\filldraw[fill=black] (90:1.2ex) -- (170:0.7ex) -- (10:0.7ex) --cycle;
	\end{tikzpicture}
}

\tikzstyle{round_rect} = [rectangle, rounded corners, minimum width=1.5cm, minimum height=0.5 cm,text centered, draw=black, fill=white!30]

\tikzstyle{rect} = [rectangle, minimum width=1.5cm, minimum height=0.8cm, text centered, text width=1.5cm, draw=black, fill=white!30]

\tikzstyle{text_node} = [rectangle, minimum width=0.5cm, minimum height=0.35cm, text centered, text width=4cm, draw=white, fill= white!30]
\resizebox{0.68\textwidth}{!}{
\begin{tikzpicture}
	
	\node (tab1) [shape=rectangle,draw=white] {
		\scriptsize
		\begin{tabular}{ccccc}
			\toprule
			
		\textbf{$f_1$: Company}&\textbf{$f_2$: Monthly}&\multirow{2}{*}{\textbf{$f_3$: Company type}}&\textbf{$f_4$: Field of }&\multirow{2}{*}{\textbf{$f_5$: Tax ID}}\\
		\textbf{name}&\textbf{ revenue}&&\textbf{ activity}&\\
		\emph{(Textual)}&\emph{(Numerical)}&\emph{(Categorical)}&\emph{(Categorical)}&\emph{(Textual)}\\
			\midrule
			
			{\color{gray}\text{UCI}$\rightarrow$} Required &{\color{gray}20$\rightarrow$}[20,22) &Large enterprise  &Real estate&T190 \\
			\midrule
			{\color{gray}{KDL $\rightarrow$}} Required&{\color{gray}21$ \rightarrow$}[20,22) &Large enterprise   &Manufacturing&T201  \\ 
			\midrule
			{\color{gray}{EoP}$\rightarrow$} Required&{\color{gray}@$\rightarrow$}Optional&Large enterprise &  Manufacturing& T200   \\ 
			\midrule
			{\color{gray}{UNI}$\rightarrow$} Required &{\color{gray}39$\rightarrow$}[39,41)&NPO&Education &{\color{gray}n/a$\rightarrow$}Optional \\

			\bottomrule
		\end{tabular}
};
\end{tikzpicture}}
 	\caption{Example of Pre-processed Historical Input Instances }
	\label{fig:preprocessing}
\end{figure}

\figurename~\ref{fig:preprocessing} shows an example of historical input instances collected from 
the data entry form presented in \figurename~\ref{fig:def}. 
During the preprocessing phase, \approach identifies meaningless values in different 
fields (e.g., ``n/a'' and ``@'')  and replaces them by the dummy value \emph{Optional}. 
For the remaining ``meaningful'' values, \approach replaces values in 
the textual field ``Company name'' to the dummy value \emph{Required};
values in the field ``Monthly revenue'' are discretized into intervals. 
In addition to historical input instances, \approach also preprocesses the input instance filled 
during the data entry session.
For example, as shown in \figurename~\ref{fig:def}, a user fills values 
\emph{Wish}, \emph{20}, \emph{NPO}, and 
\emph{Education} in fields ``Company name'', ``Monthly revenue'', ``Company type'',
and ``Field of activity'', respectively.
\approach will replace the value filled in the field ``Company name'' to ``Required'', 
since it is a meaningful 
value.
\approach also maps the value in the field ``Monthly revenue'' into the interval \emph{[20, 22)}.

\subsection{Model Building}
\label{sec:model-building}

The \emph{model building} phase aims to learn the completeness requirement dependencies between different fields 
from training input instances related to a data entry form. 

During the data entry session, 
we consider the filled fields as features to predict the completeness requirement of the target field (i.e., optional or required).
However, in our previous work~\cite{belgacem2022machine} we have shown that
in an extreme scenario, users could follow any arbitrary order to fill the form,
resulting in a large set of feature-target combinations.
For example, given a data entry form with $n$ fields, when we consider one of the fields as the target, 
we can get a total number of up to $2^{n-1}-1$ 
feature (i.e., filled fields) combinations.
Based on the assumption of identical features and targets ~\cite{dekel2010learning} to train and test a machine learning model,
a model needs to be trained on each feature-target combination,
which would lead to training an impractical large number of models. 

To deal with this problem, we select BNs as the machine learning models to 
capture the completeness requirement dependencies between filled fields and 
the target field,
without training models on specific combinations of fields.
As already discussed  in our 
 previous work~\cite{belgacem2022machine}, 
the reason is that
BNs can infer the value of a target field using only information
in the filled fields and the related PGM (see
section~\ref{sec:bayesian-networks}); BNs automatically deal with the
missing conditionally independent variables (i.e.,  unfilled
fields).

In this work,
\approach learns the BN structure representing the completeness requirement dependencies 
from training input instances.
Each field in the data entry form represents a node (random variable)
in the BN structure; the edges between different nodes are the dependencies between different fields. 
In order to construct the optimal network structure,
BN performs a search-based optimization 
based on the conditional probabilities of the fields and a fitness function. 
As in our previous work~\cite{belgacem2022machine}, 
we use hill climbing as the optimizer to learn the BN structure
with a fitness function based on  BIC~\cite{raftery1995bayesian}.

\begin{algorithm}[tbp!]
	\footnotesize
	\caption{Model Building}
	\label{alg:LACQUER-training}
	\KwIn{Set of preprocessed historical input instances $I^F(t)_{\mathit{train}}^\prime$ for training}
	\KwOut{Dictionary of probabilistic graphical models $\mathcal{M}$}
	
	$\mathcal{M}\gets$empty dictionary\;
	List of fields $\mathit{fields}\gets \mathit{getFields}(I^F(t)_{\mathit{train}}^\prime) $\; \label{line:getfields}
	\ForEach{field $f_i \in \mathit{fields}$}
	{
		Temporary training set $\mathit{train}_{f_{i}}\gets \mathit{getDataSetForField}(I^F(t)_{\mathit{train}}^\prime, f_i)$\;\label{line:temp_training_set}
		Oversampled Training set $\mathit{train}_{f_{i}}^{\mathit{oversample}}\gets \mathit{SMOTE}(\mathit{train}_{f_{i}}, f_i)$\;\label{line:smote}
		Model $M_i\gets \mathit{trainBayesianNetwork}(\mathit{train}_{f_{i}}^{\mathit{oversample}})$\; \label{line:train_on_oversampled_dt}
		$\mathcal{M}[f_i  ]\gets M_i$\label{line:save_M}
	}	\label{line:train-e-fields}
	\KwRet{$\mathcal{M}$}\;
\end{algorithm}

\begin{figure*}[tbp!]
	\centering
	
\newcommand{\tikzmark}[2]{
	\tikz[overlay,remember picture,baseline] 
	\node [anchor=base] (#1) {$#2$};}

\newcommand{\DrawVLine}[3][]{\begin{tikzpicture}[overlay,remember picture]
		\draw[shorten <=0.3ex, #1] (#2.north) -- (#3.south);
	\end{tikzpicture}
}

\newcommand{\DrawCircle}[1][]{\begin{tikzpicture}[overlay,remember picture]
		\draw[black,fill=black] (0,0.5ex) circle (0.5ex);
	\end{tikzpicture}
}

\newcommand{\DrawRec}[1][]{\begin{tikzpicture}[overlay,remember picture]
		\draw[black,fill=black] (-0.5ex,0) rectangle ++(1ex, 1ex);
	\end{tikzpicture}
}

\newcommand{\DrawTriangle}[1][]{\begin{tikzpicture}[overlay,remember picture]
		\filldraw[fill=black] (90:1.2ex) -- (170:0.7ex) -- (10:0.7ex) --cycle;
	\end{tikzpicture}
}

\tikzstyle{round_rect} = [rectangle, rounded corners, minimum width=1.5cm, minimum height=0.5 cm,text centered, draw=black, fill=white!30]

\tikzstyle{rect} = [rectangle, minimum width=1.5cm, minimum height=0.8cm, text centered, text width=1.5cm, draw=black, fill=white!30]

\tikzstyle{text_node} = [rectangle, minimum width=0.5cm, minimum height=0.35cm, text centered, text width=4cm, draw=white, fill= white!30]

	\resizebox{1\textwidth}{!}{
\begin{tikzpicture}

	\node (tab1) [shape=rectangle,draw=white] {
		\scriptsize
		\begin{tabular}{ccccc}
			\toprule
		
		\textbf{$f_2$: Monthly}&\multirow{2}{*}{\textbf{$f_3$: Company type}}&\textbf{$f_4$: Field of }&\multirow{2}{*}{\textbf{$f_5$: Tax ID}}\\
		\textbf{ revenue}&&\textbf{ activity}&\\
		\emph{(Numerical)}&\emph{(Categorical)}&\emph{(Categorical)}&\emph{(Textual)}\\
		\midrule
		
		$[20,22)$ &Large enterprise&Real estate&T190 \\
		\midrule
		$[20,22)$ &Large enterprise &Manufacturing& T201 \\ 
		\midrule
		Optional&Large enterprise  &Manufacturing & T200\\ 
		\midrule
		$[39,41)$&NPO& Education& Optional\\ 
		\bottomrule
	
		\end{tabular}
	};

\node(init_hist) [text_node, below=of tab1, text width=8cm,yshift=1.0cm]{\textbf{Preprocessed historical input instances}};
\node (tab2) [shape=rectangle,draw=white, below = of  init_hist, yshift=0.5cm] {
	\scriptsize
	\begin{tabular}{ccccc}
		\toprule
		\textbf{$f_2$: Monthly}&\multirow{2}{*}{\textbf{$f_3$: Company type}}&\textbf{$f_4$: Field of }&\multirow{2}{*}{\textbf{$f_5$: Tax ID}}\\
		\textbf{ revenue}&&\textbf{ activity}&\\
		\emph{(Numerical)}&\emph{(Categorical)}&\emph{(Categorical)}&\emph{(Textual)}\\
		\midrule
		
		$[20,22)$ &Large enterprise&Real estate&\color{red}\text{Required}\\
		\midrule
		$[20,22)$ & Large enterprise&Manufacturing& \color{red}\text{Required} \\ 
		\midrule
		Optional&Large enterprise&Manufacturing& \color{red}\text{Required}\\ 
		\midrule
				$[39,41)$&NPO &Education& \color{red}\text{Optional}\\ 
		\bottomrule
		
	\end{tabular}
};
\node(train_for_target) [text_node, below=of tab2,text width=8cm,yshift=1.0cm]{\textbf{Temporary training set for target ``Tax ID''}};

\node (tab3) [shape=rectangle,draw=white, right = of  tab2, yshift= 0.45cm] {
	\scriptsize
	\begin{tabular}{ccccc}
		\toprule
		
	\textbf{$f_2$: Monthly}&\multirow{2}{*}{\textbf{$f_3$: Company type}}&\textbf{$f_4$: Field of }&\multirow{2}{*}{\textbf{$f_5$: Tax ID}}\\
	\textbf{ revenue}&&\textbf{ activity}&\\
	\emph{(Numerical)}&\emph{(Categorical)}&\emph{(Categorical)}&\emph{(Textual)}\\
	\midrule
		
		$[20,22)$ &Large enterprise&Real estate&\color{red}\text{Required}\\
		\midrule
		$[20,22)$ &Large enterprise&Manufacturing& \color{red}\text{Required} \\ 
		\midrule
		Optional&Large enterprise &Manufacturing& \color{red}\text{Required}\\ 
		\midrule
		$[39,41)$& NPO &Education& \color{red}\text{Optional}\\ 
		\midrule
		$[39,41)$&NPO&Education& \color{red}\text{Optional}\\ 
		\midrule
		$[39,41)$&NPO &Education& \color{red}\text{Optional}\\ 
		\bottomrule
	\end{tabular}
};

\node(overampled_train) [text_node, below=of tab3,text width=8cm,yshift=1.0cm]{\textbf{Oversampled training set}};

\begin{scope}[transform canvas={xshift=0em},node distance=2mm]
	\node (f3) [round_rect, minimum width=1cm,above = of tab3,yshift= 2.34cm] {$f_3$};
	\node (f4) [round_rect,minimum width=1cm, below= of f3] {$f_4$};
	\node (f2) [round_rect, minimum width=1cm,below left= of f4,xshift=0.6cm] {$f_2$};
	\node (f5) [round_rect, minimum width=1cm,below right= of f4,xshift=-0.6cm] {$f_5$};
	\draw [arrow,->] (f3) -- (f4);
	\draw [arrow,->] (f2) -- (f4);
	\draw [arrow,->] (f4) -- (f5);
\end{scope}
\node (big_box)[draw=black, fit={(f3)(f4)(f2)(f5)},minimum width=2.4cm,minimum height=2.0cm,line width=0.3mm, above = of tab3, yshift= -0.1cm]{};

\node(model_for_target) [text_node, below=of big_box,text width=8cm,yshift=0.9cm]{\textbf{The BN model for ``Tax ID''}};
\draw[arrow,->] (init_hist)--(tab2);
\draw[arrow,->] (tab2)++(4,0.35)-- ++(0.8,0)(tab3);
\draw[arrow,->] (tab3)--(model_for_target);

\node[shape=circle, draw, fill=gray, opacity=.2, text opacity=1, inner sep=0.5pt, left=of tab1, xshift=1cm, yshift=0.1cm] (charA) {A};

\node[shape=circle, draw, fill=gray, opacity=.2, text opacity=1, inner sep=0.5pt, left=of tab2, xshift=1cm, yshift=0.1cm] (charB) {B};

\node[shape=circle, draw, fill=gray, opacity=.2, text opacity=1, inner sep=0.5pt, left=of tab3, xshift=1cm, yshift=0.4cm] (charC) {C};

\node[shape=circle, draw, fill=gray, opacity=.2, text opacity=1, inner sep=0.5pt, left=of big_box, xshift=0.8cm, yshift=0.4cm] (charD) {D};

\end{tikzpicture}
} 	\caption{Workflow of the Model Building Phase}
	\label{fig:lacquer-modelbuilding}
\end{figure*}
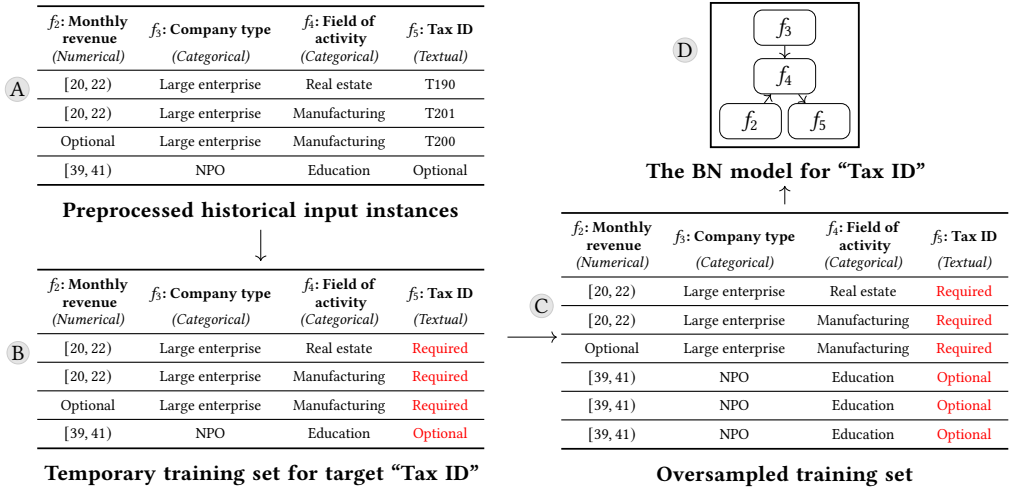

Algorithm~\ref{alg:LACQUER-training} illustrates the main steps of this phase.
\approach takes as input a set of preprocessed historical input instances  $I^F(t)_{\mathit{train}}^\prime$
for training and learning the completeness requirement dependencies 
(e.g., the input instances in block \myblock{A} of \figurename~\ref{fig:lacquer-modelbuilding}).
Initially, for each field $f_i$ in the list of fields extracted from $I^F(t)_{\mathit{train}}^\prime$ (line~\ref{line:getfields}), 
we create a temporary training set where we consider the field $f_i$ as the target (line~\ref{line:temp_training_set}).
Since we aim to predict whether the target field is required or optional during form filling,
in the temporary training set, we keep the value ``Optional'' in the target field  $f_i$ and label other values as ``Required'' (block \myblock{B} in \figurename~\ref{fig:lacquer-modelbuilding}).
These two values are the classes according to which to predict $f_i$.

However, we may not train effective classification models directly on this temporary training set.
This is caused by the imbalanced nature of input instances for different classes.
Users commonly enter correct and meaningful values during form filling.
They only fill meaningless values in certain cases.
As a result, the number of input instances having meaningless values (i.e., in the ``Optional'' class) is usually smaller than the number of input instances in the ``Required'' class.
This can make the learning process inaccurate~\cite{johnson2019survey},
since machine learning models may consider the minority class as noise~\cite{roshan2020improvement}. 
The trained models could also over-classify the majority class due to its increased prior probability~\cite{johnson2019survey}.
For example in block \myblock{B} of \figurename~\ref{fig:lacquer-modelbuilding}, 
considering that the column ``$f_5$: Tax ID'' is the current target, 
the number of instances in class ``Required'' is three,
which is  higher than the single instance in class
``Optional''. 
If we train a model on such imbalanced dataset, 
it might be difficult to learn the conditions (or dependencies) to relax this field as optional due to the small number of ``Optional'' instances. 

To solve this problem,
we apply SMOTE (line~\ref{line:smote}) on the temporary training set $\mathit{train}_{f_{i}}$
to generate an oversampled training set $\mathit{train}_{f_{i}}^{\mathit{oversample}}$ (as shown in block \myblock{C} in \figurename~\ref{fig:lacquer-modelbuilding}). 
After oversampling, both classes have the same number of input instances.
We train a BN
model $M_i$ based on the oversampled training set for the target field $f_i$ (line~\ref{line:train_on_oversampled_dt}).
For example, block \myblock{D} in \figurename~\ref{fig:lacquer-modelbuilding} 
represents the model built for the target field ``Tax ID''.
Following this step, we can obtain a BN model for each field.
We save all the BN models in the dictionary $\mathcal{M}$ (line~\ref{line:save_M}), 
where the key represents the name of the field
and the value is the corresponding trained BN model. 
The output of Algorithm~\ref{alg:LACQUER-training} is the dictionary $\mathcal{M}$.

\subsubsection*{Application to the running example}

Given the preprocessed training input instances  shown in block 
\myblock{A} in \figurename~\ref{fig:lacquer-modelbuilding}, 
\approach creates a temporary training set for each target (e.g., the field ``Tax ID''),
where \approach replaces
the meaningful and meaningless values of the target field to \emph{Required} and \emph{Optional}, respectively (in block \myblock{B}).
The temporary training set is oversampled using SMOTE to create a balanced training set
where the number of instances of both Required and Optional classes is the same (block \myblock{C} of \figurename~\ref{fig:lacquer-modelbuilding}).
This oversampled training set is used to train a BN model for the target field ``Tax ID''. 
An example of the trained BN model is presented in block \myblock{D} in \figurename~\ref{fig:lacquer-modelbuilding}. 
After the model building phase, \approach outputs a model for each target. 
For the example of training input instances  related to
\figurename~\ref{fig:def}, \approach returns five distinct models where each model captures the completeness requirement dependencies for a given target.

 \subsection{Form Filling Relaxation}
\begin{figure*}[tb]
	\centering
	
\tikzstyle{round_rect} = [rectangle, rounded corners, minimum width=1.5cm, minimum height=0.5 cm,text centered, draw=black, fill=white!30]
\tikzstyle{rect} = [rectangle, minimum width=1.5cm,  text centered, text width=1.5cm, draw=black, fill=white!30]
\tikzstyle{text_node} = [rectangle, minimum width=0.5cm, text centered, text width=4cm, draw=white, fill= white!30]
\tikzstyle{text_model} = [rectangle, minimum width=0.8cm, text centered, text width=0.8cm, draw=white, fill= white!30]

\tikzstyle{list_box} = [rect,align=left ,minimum width=2.39cm,text width=3.19cm, minimum height= 0.5cm]

\resizebox{1\textwidth}{!}{
	\begin{tikzpicture} [node distance=0.3mm, >=latex]
	
	\node (input_val) [rect, text
        width=2.5cm,xshift=0.5cm]{\textbf{Filled fields}\\   
           $f_2$:Monthly revenue=\emph{[20, 22]}\\$f_3$:Company type =\emph{NPO}\\ $f_4$:Field of activity =\emph{Education}\\$ $\\ \textbf{Target}\\ $f_5$:Tax ID= \emph{?}};		
	\node (current_input)[text_node, below = of input_val, yshift=-0.2cm]{Pre-processed \\current input};

	\begin{scope}[transform canvas={xshift=1em}, node distance=2mm]
		\node (m0f4) [round_rect, minimum width=0.2cm, minimum height=0.08cm, right =of input_val.north,xshift=2.8cm, yshift=-0.6cm] { };
		\node (m0f3) [round_rect, minimum width=0.2cm, minimum height=0.08cm, above=of m0f4] { };
		\node (m0f2) [round_rect, minimum width=0.2cm, minimum height=0.08cm, below left= of m0f4, xshift=0.6cm] { };
		\node (m0f5) [round_rect, minimum width=0.2cm, minimum height=0.08cm, below right= of m0f4, xshift=-0.6cm] { };
		\draw [arrow,->] (m0f3) -- (m0f4);
		\draw [arrow,->] (m0f5) -- (m0f4);
		\draw [arrow,->] (m0f4) -- (m0f2);
	\end{scope}

	\node (cluster1) [rect, draw=none, right =of input_val.north, minimum height= 0.4cm, yshift=-1.8cm, xshift=2.2cm, text width=2.2cm]{ Model for target ``Tax ID''};

	\node (textpoints) [text_node, below= of cluster1, text width=0cm, xshift=-0.2cm, yshift= -0.1cm] {....};
	
	\begin{scope}[transform canvas={xshift=1.5em}, node distance=2mm,yshift=1cm]
		\node (mkf4) [round_rect, minimum width=0.2cm, minimum height=0.08cm, below=of textpoints, xshift=-0.1cm, yshift=-0.2cm,] { };
		\node (mkf3) [round_rect, minimum width=0.2cm, minimum height=0.08cm, above left=of mkf4, , yshift=-0.1cm] { };
		\node (mkf2) [round_rect, minimum width=0.2cm, minimum height=0.08cm, below left= of mkf4, yshift=0.1cm] { };
		\draw [arrow,->] (mkf3) -- (mkf4);
		\draw [arrow,->] (mkf2) -- (mkf4);
	\end{scope}
	\node (cluster3) [rect, draw=none, below=of textpoints, xshift=0.25cm, yshift=-0.9cm, minimum height= 0.4cm, text width=2.2cm]{Model for target ``Company type''};
	
	\node (model_tax_id)[draw=none, fit={(m0f4)(m0f3)(m0f2)(m0f5)},minimum width=1.8cm,minimum height=1cm]{};	
	\node (model_field_of_activity)[draw=none, fit={(mkf4)(mkf3)(mkf2)},minimum width=1.4cm,minimum height=1cm]{};	
	
	\node (anchor1) [text_node, right= of cluster1, yshift= 1.3cm, xshift=-0.56cm, text width=0.1cm] {};
	\node (anchor2) [text_node, right= of cluster3, yshift= -0.0cm, xshift=-0.65cm, text width=0.0cm] {};
	
	\draw [decorate, xshift=1cm, decoration={brace, aspect=.5, amplitude=10pt, raise=4pt}] (anchor1.north east) -- (anchor2.south east)  node [black] { };
	
	\begin{scope}[transform canvas={xshift=0em}, node distance=3mm]
		\node (f4) [round_rect, minimum width=1cm, right= of textpoints, xshift=1.8cm, yshift=-0.5cm] {$f_4$};
		\node (f3) [round_rect, minimum width=1cm, above= of f4] {$f_3$};
		\node (f2) [round_rect, minimum width=1cm, below left= of f4,xshift=0.7cm] {$f_2$};
		\node (f5) [round_rect, minimum width=1cm, below right= of f4,xshift=-0.7cm] {$f_5$=?};
		\draw [arrow,->] (f3) -- (f4);
		\draw [arrow,->] (f2) -- (f4);
		\draw [arrow,->] (f4) -- (f5);
	\end{scope}

	\node (sel_explain) [text_node, above= of f3, yshift= 0cm, text width=2.0cm] {\\Model for \\``Tax ID'' \\ is selected};
	
	\node (big_box_selection)[draw=black, fit={(model_field_of_activity) (model_tax_id)(cluster1)(cluster3)(f2)(f3)(f4)(f5)},minimum width=1.5cm,minimum height=2.5cm,yshift=-0.23cm]{};
	
	\draw [arrow,->] (input_val.east) ++(0, -0.01)--++(0.7, 0) (big_box_selection);
	
	\node (sel_explain2) [text_node, below= of big_box_selection, yshift= 0cm, text width=3.5cm] {\myblock{A} Model selection};
	
	\node (table) [shape=rectangle,draw=white, right = of f4, yshift=0.45cm, xshift=1.5cm] {
		\begin{tabular}{p{1.6cm}c}
		\toprule
		\textbf{Class}& \textbf{Prob.}\\
		\midrule
		Optional& 0.80  \\
		Required & 0.20  \\
		\bottomrule
		\end{tabular}
	};

	\node(text_distribution) [text_node, above = of  table,minimum width=0.8cm,text width=3.5cm]{Probability distribution};

	\node (text_sug) [text_node, right=of table, yshift=-2mm, xshift= 0.3cm]{Suggestion of \\ ``Tax ID'' is \emph{Optional}};
	\draw [arrow,->] (table.east) ++(0, -0.24)-- ++(0.9, 0)(text_sug);

	\node (big_box_suggestion)[draw=none, fit={(table)},minimum width=1.4cm,minimum height=2.5cm]{};

	\draw [arrow,->] (big_box_selection.east) ++(0, 0.02)--++(0.8, 0) (big_box_suggestion);

	\node (check_prob)[ text_node, below= of big_box_suggestion, text width=3.5cm, minimum height=0.5cm, draw=none, yshift=0cm, xshift=2.3cm] {Prob.=0.8 $>$ $\theta_p$=0.70 \\ \emph{checkProb}=\emph{true}};
	
	\node (big_box_endorser)[draw=black, fit={(check_prob)},minimum width=1.4cm,minimum height=1.0cm]{};
	
	\node (endorser_explain) [text_node, below= of big_box_endorser, yshift= -0.4
	cm, text width=2.5cm] {\myblock{B} Endorsing};
	
	\draw [arrow, ->{Circle}] (big_box_endorser.north) ++(-0.1,0)--++(0,0.99) (big_box_suggestion);

	\end{tikzpicture}
} 	\caption{Workflow for Form Relaxation Phase}
	\label{fig:selection-filter}
\end{figure*}
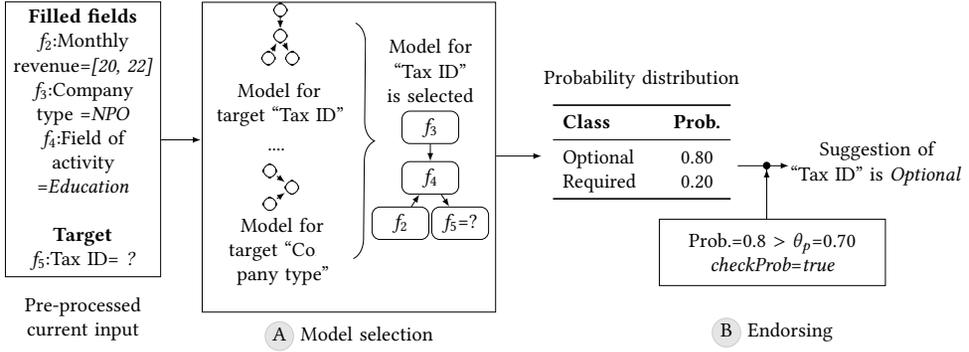

\begin{algorithm}[tbp]
	\footnotesize
	\caption{Form Filling Relaxation}
	\label{alg:relaxation}
	\KwIn{Dictionary of Models $\mathcal{M}=\{f_1:M_1, \dots, f_k:M_k\}$ \newline
		Set of 	Filled fields $C_t=\{\langle f^c_1, v^c_1 \rangle, \dots, \langle f^c_m, v^c_m\rangle\}$ \newline
		Target field $f_p$ \newline
		Threshold $\theta_p$
	}
	\KwOut{A Boolean $\mathit{checkOP}_p$, representing the decision to set field $f_p$ to optional}
	
	Set $C_{t}' \gets \mathit{getPreprocessedData}(C_{t} )$ \;\label{line:pre-processing}
	Model $M_{p} \gets \mathcal{M}[f_p]$\;\label{line:model-selection}

	List of pairs of completeness requirements and probability distribution 
	$\langle \mathit{cr}, \mathit{pr}\rangle_{\mathit{list}} =
	\mathit{predictCR}(M_{\mathit{p}}, C_{t}' , f_p)$\; \label{line:prediction}
	
	 Top-ranked pair 
	 $\langle \mathit{cr_{\mathit{top}}}, \mathit{pr_{\mathit{top}}}\rangle =
	 \mathit{getTopRanked}(\langle \mathit{cr}, \mathit{pr}\rangle_{\mathit{list}})$\; \label{line:rank prediction}
	 Boolean  $\mathit{checkOP}_p \gets \mathit{isOptional}(\mathit{cr_{\mathit{top}}}
	)$\; \label{line:isopt}
	\If{$ \mathit{checkOP}_p $} 
	{	Boolean $\mathit{checkProb} \gets (pr_\mathit{top} < \theta_p)$\; \label{line:check-prob}   
		\If{$ \mathit{checkProb} $}
		  {
			{$\mathit{checkOP}_p \gets \mathit{``false}"$;\label{line:consider-it-req}}
	     }
    }
	
	\KwRet{$\mathit{checkOP}_p$}\;
\end{algorithm}

The form filling relaxation phase is an online phase that occurs during the data entry session. 
In this phase, \approach selects the model $M_{p}$ $\in \mathcal{M}$ corresponding to the target field $f_p$.
This model is then used to predict the completeness requirement of the target $f_p$
based on the filled fields $C_{t}$.
The main steps are shown in Algorithm~\ref{alg:relaxation}.

The inputs of the algorithm are the dictionary of trained models $\mathcal{M}$, 
the set $C_{t}$ representing the filled fields during the entry session and their values, 
the target field $f_p$, and the endorsing threshold $\theta_p$ for $f_p$.
The algorithm starts by applying the preprocessing techniques outlined in \S~\ref{sec:pre-processing}
to the set of the filled fields in $C_{t}$ in order to obtain a new set of preprocessed
filled fields $C_{t}'$ (line~\ref{line:pre-processing}).
\approach then selects the model $M_p$ from $\mathcal{M}$ (line~\ref{line:model-selection}),
since this model is trained for the target field $f_p$ based on the oversampled data
with a balanced number of instances for each class of $f_p$.
With the selected model, 
\approach predicts the completeness requirement for $f_p$ (line~\ref{line:prediction}) 
and gets the top-ranked completeness requirement based
on the prediction probability (line~\ref{line:rank prediction}).

\subsubsection*{Endorsing}
During the data entry session,
values in filled fields do not always provide enough knowledge 
for the model to accurately predict the completeness requirement of a given target field.
This happens because when training BN models,
there may not be enough information in the training input instances to learn the dependencies among some 
fields with specific values.

However, in the context of form filling relaxation, it is important to provide accurate completeness requirement suggestions.
On the one hand, wrongly predicting optional fields as required adds more constraints to the data entry form; users 
will be obliged to fill fields with meaningless values.
On the other hand, wrongly predicting a required field as optional can result in missing information.
In order to prevent this situation, \approach includes a heuristic-based endorser module that
decides if the predicted completeness requirement is correct or not.
Since our main goal is to relax the completeness requirement by predicting when a required field
should be optional,
we mainly use the endorser to endorse the prediction where the target field is predicted as ``Optional''.
If the prediction is endorsed, we set the field to ``Optional''; otherwise, we use its previous setting (``Required'').
 
Specifically, \approach checks if the top-ranked completeness requirement is
equal to ``Optional''; it saves the result in the Boolean flag $\mathit{checkOP}_p$ (line~\ref{line:isopt}).
 If the value of  $\mathit{checkOP}_p$  evaluates to \emph{true},
\approach analyses the probability distribution of the predicted completeness
requirement of the target field
since it reflects whether \approach has enough ``confidence'' in the prediction based on current information.
We check if the probability for the field to be ``Optional'' is
lower than a threshold $\theta_p$ for target $f_p$
(line~\ref{line:check-prob}), saving the result in the Boolean flag \emph{checkProb}. 
If the value of \emph{checkProb} evaluates to \emph{true}, 
we change the value of the Boolean flag $\mathit{checkOP}_p$ to \emph{false} (line~\ref{line:consider-it-req}) since it implies the model does not have enough ``confidence'' for variable inference and prediction; otherwise, 
\approach keeps the prediction as ``Optional''.
The threshold $\theta$ is automatically determined; 
its value differs for different targets (as discussed in \S~\ref{sec:tuning}).
We use different threshold values because the prediction is done by models 
trained on different data and the variance of the data can have a significant effect on prediction accuracy~\cite{wu2021sampling}.

\subsubsection*{Application to the running example}

\figurename~\ref{fig:selection-filter} shows the process of predicting the completeness requirement of the field 
``Tax ID'' based on the input values in
 \figurename~\ref{fig:def}.
 \approach first selects the model related to the current target for prediction (block \myblock{A} in \figurename~\ref{fig:selection-filter}).
 Let us assume that, based on the BN variable
 inference, 
 \approach predicts that field ``Tax ID'' has a probability of 0.80 to be Optional.
 Since the top predicted value is Optional, \approach activates the endorser module (block \myblock{B} in  \figurename~\ref{fig:selection-filter})
 to decide whether the level of confidence is acceptable. 
For example, let us assume  the automatically decided threshold value for field ``Tax ID'' is 0.70 (i.e., $\theta_{\mathit{tax ID}}$=0.70). 
Since the probability value of the ``Optional'' class (0.80) is higher than
this threshold, the Boolean flag
$\mathit{checkOP}_\mathit{TaxID}$ remains true.
\approach decides to set the field ``Tax ID'' to Optional.

\subsection{Endorser Threshold Determination}
\label{sec:tuning}

\begin{algorithm}[tbp]
	\footnotesize
	\caption{Endorser Threshold Determination}
	\label{alg:LACQUER-tuning}
	\KwIn{
	   Set of pre-processed historical input instances $I^F(t)_{\mathit{tune}}^\prime$ for tuning \newline
	   Dictionary of Models $\mathcal{M}=\{f_0: M_0, f_1:M_1, \dots, f_k:M_k\}$
	}
	\KwOut{Dictionary of thresholds $\theta$}
	$\theta \gets$empty dict\;
	List of fields $\mathit{fields}\gets \mathit{getFields}(I^F(t)_{\mathit{tune}}^\prime) $\; \label{line:tuning-getfields}
	\ForEach{field $f_i \in \mathit{fields}$}
	{
		$\mathit{temp_{th}}\gets$empty dictionary\;
		$I^F(t)_{\mathit{tune}_i}^\prime = \mathit{getDataSetForField}(I^F(t)_{\mathit{tune}}^\prime, f_i)$\;\label{line:tuning-temp-data}
		Model $M_{i} \gets \mathcal{M}[f_i]$\;\label{line:tuning-model-selection}
		
		\For{n= 0 to 1 (step 0.05) }
		{
			${\mathit{predictedCRAll}} =
			\mathit{predictCRAllInstances}( M_{\mathit{i}}, I^F(t)_{\mathit{tune}_i}^\prime, n)$\; \label{line:tuning-prediction-details}
			${\mathit{score}} =
			\mathit{evaluate}(I^F(t)_{\mathit{tune}_i}^\prime, \mathit{predictedCRAll})$\; \label{line:tuning-compute-score}
			${\mathit{temp_{th}} [n]} = \mathit{score}$\; \label{line:tuning-save-score}
		}
      ${\mathit{\theta[i]}} =
      \mathit{getBestScore}(\mathit{temp_{th}})$\; \label{line:tuning-best-th}
	}	
	
	\KwRet{$\theta$}\;
\end{algorithm}

We automatically determine the value of the threshold
in the endorser module for each target.
This step occurs offline and assumes that the models in $\mathcal{M}$
built during the model building phase are available.
The threshold $\theta_i$ for the target field $i$
is determined with the set of preprocessed tuning input instances.
The basic idea is that 
for each historical input instance in this subset,
we consider all fields except field $i$ to be filled 
and use the model trained for field $i$ to predict its completeness requirement with different values of $\theta_i$.
We determine the value of $\theta_i$
based on the value that achieves the highest prediction accuracy 
on tuning input instances. 

The main steps are shown in Algorithm~\ref{alg:LACQUER-tuning}.
The algorithm takes as input the set of preprocessed historical input instances for tuning $I^F(t)_{\mathit{tune}}^\prime$ and
the trained models $\mathcal{M}$.
For each field $f_i$ in the list of fields extracted from $I^F(t)_{\mathit{tune}}^\prime$ (line~\ref{line:tuning-getfields}),
we generate a temporary dataset $I^F(t)_{\mathit{tune}_i}^\prime$ where the value of field $f_i$ is transformed into ``Optional'' and ``Required'' 
using the method presented in \figurename~\ref{fig:lacquer-modelbuilding}(B) (line~\ref{line:tuning-temp-data}).
We select the model corresponding to $f_i$ from $\mathcal{M}$ (line~\ref{line:tuning-model-selection}) and
use the selected model to predict the completeness requirement of field $f_i$ 
based on the values of other fields in $I^F(t)_{\mathit{tune}_i}^\prime$ (line~\ref{line:tuning-prediction-details}).
While predicting, we try different thresholds, 
varying from 0 to 1 with a step equal to 0.05.
For each threshold value,
we compare the predicted completeness requirement with the actual completeness requirement of field $f_i$ in each input instance of $I^F(t)_{\mathit{tune}_i}^\prime$ to calculate the prediction accuracy (line~\ref{line:tuning-compute-score}). 
\approach selects the value of $\theta_i$
that achieves the highest prediction accuracy value in $I^F(t)_{\mathit{tune}_i}^\prime$
as the threshold for $f_i$ (line~\ref{line:tuning-best-th}).
The algorithm ends by returning a dictionary containing the thresholds of all fields.

 \section{Evaluation}
\label{sec:evaluation}

In this section, we report on the evaluation of our approach
for automated completeness requirement relaxation. 
First, we assess the overall accuracy of \approach when predicting the 
completeness requirement of fields in data entry forms, 
and compare it with state-of-the-art baselines.
We then assess the performance of \approach, in terms of training time and prediction time, for practical applications.
Last, we perform an ablation study to evaluate how the use of SMOTE (in the model building phase)
 and the heuristic-based endorser (in the form filling relaxation phase) 
 affects the accuracy of \approach.

More specifically, we evaluated \approach by answering the following
research questions (RQs):
\begin{compactenum}[RQ1]
\item \emph{Can \approach provide accurate predictions for completeness requirement relaxation, 
	and how does it compare with baseline algorithms?}
\item \emph{Is the performance of \approach, in terms of training 
  and prediction time, suitable for practical applications?}
\item \emph{What is the impact of using SMOTE and the
    heuristic-based endorser on the effectiveness of \approach?}
\end{compactenum}

\subsection{Dataset and Settings}
\label{sec:dataset_and_setting}

\begin{table}[tbp]
  \caption{Information about the Fields in the Datasets}
	\centering
	\scriptsize
	\begin{filecontents*}{data/dataset-info.csv}
Dataset; 			     Fields; 				 Instances; 	number_required;			Categorical
\multirow{2}{*}{NCBI};	 \multirow{2}{*}{26};    \multirow{2}{*}{235538}; \multirow{2}{*}{6} ;     sample-name(0), tissue(0.130), isolate(0.351), sex(0.351)

					 ;						;						   ;	; biomaterial-provider(0.1), age(0.543)				

\multirow{7}{*}{PEIS} ;	 \multirow{7}{*}{33};	 \multirow{7}{*}{73082};	\multirow{7}{*}{19}; legal name(0), contact name(0), first name(0.113),
					 ;						;							;	; place of birth(0.127), native country(0.127), status(0),									
					 ;						;							;	; year of study(0.94), function(0), employer name(0.35),							
					 ;						;							;	; name of school/university(0.84), type of contract(0), 
				         ;						;							;;	contract start date(0.668), date of end of contract(0.974),							
					;						;							;	;field of activity (0), code NACE(0.123), primary activity(0),
					;						;							;	; country of activity(0),percentage of activity(0)
\end{filecontents*}
\pgfplotstabletypeset[
	every head row/.style={
		before row={\toprule
			\multirow{2}{*}{Dataset}& \# of & \# of & \# of&Name of required fields\\ 
		},
		after row=\midrule,
	},
	every last row/.style={
		after row=\bottomrule},
	every nth row={8[+2]}{before row=\midrule},
	columns/Dataset/.style ={column name=}, 
	columns/Fields/.style ={column name=fields}, 
	columns/Instances/.style ={column name=instances},
	columns/number_required/.style= {column name= required fields},
	columns/Categorical/.style={column name=(\% of missing and meaningless values)},	
col sep=semicolon,
	string type,
	]{data/dataset-info.csv}
	\label{tab:dataset-info}
\end{table}

\subsubsection*{Datasets}
We selected the datasets used for the evaluation of \approach according to 
	the following criteria:
\begin{inparaenum}[(1)]
	\item data should be collected from a real
          data entry form;
          \item the form fields should have  different completeness requirements (i.e., required and  optional); and 
	\item the data entry form should have obsolete required
          fields, 
where users could use meaningless values to pass the validation
checks.
      \end{inparaenum}

We identified two datasets meeting these criteria: one publicly available in the biomedical domain (dubbed NCBI) 
and another proprietary dataset, extracted from a production-grade enterprise information system, provided by our industrial partner (dubbed PEIS).
Each dataset consists of data collected from \emph{one} real-world
  data entry form.

Other datasets used in related work on adaptive data entry forms
	(see also section~\ref{sec:related_work}) were either not mentioned ~\cite{frank1998adaptive, thistlewaite1996active},
	unavailable~\cite{bohoj2011adapforms}, or
	confidential~\cite{stromsted2018dynamic}.
	In addition, we also analyzed datasets from surveys conducted in
	countries with transparency policies (e.g., the USA ``National Survey
	on Drug Use and Health''~\cite{Survey}). However, these surveys do not
	contain a detailed specification defining the completeness requirement
	of each field  and thus the corresponding dataset does not meet our selection criterion \#2.

Both datasets are represented by a data table where each row corresponds to an input instance filled by a user
and each column represents a specific field in the data entry form;
 an input instance represents all the field values as submitted
 by a user.

The NCBI dataset is composed of metadata for diverse types of
biological samples from multiple species~\cite{barrett2012bioproject};
it has been used in previous work on automated form filling~\cite{martinez2019using,belgacem2022machine}.
This dataset provides the design of
the corresponding data entry form in the CEDAR
workbench~\cite{goncalves17:_cedar_workb} with the list of
completeness requirements for different fields.
Following the evaluation methodology described
in previous work~\cite{martinez2019using}, we considered a specific subset from the NCBI
dataset related to the species ``Homo sapiens'' for evaluation.
We downloaded the dataset  
from the official NCBI website\footnote{\url{https://ftp.ncbi.nlm.nih.gov/biosample/}}.

As shown in Table~\ref{tab:dataset-info}, the NCBI dataset contains \num{235538} instances\footnote{The number of instances 
	is different from that indicated in our previous work~\cite{belgacem2022machine} since the preprocessing step in that work retained only instances with at least three fields being filled. 
	In contrast, in this work we keep fields with missing values to analyze completeness requirements.} 
and  has 26 fields, six of which are required.
These six fields are always required and are not subject to any additional conditions.
We identify the meaningless values in the required fields using the strategy presented in section~\ref{sec:pre-processing}, 
i.e., mapping the actual value in the data  
with the dictionary of meaningless values obtained from the domain knowledge. 
In Table~\ref{tab:dataset-info}, next to each field we indicate 
 the ratio of instances having missing or meaningless values. 
The ratio of meaningless values\footnote{Required fields in
  the NCBI dataset have no missing values since they are always
  required.} varies from 0.1 (for \emph{biomaterial-provider}) to 0.543 (for
\emph{age}).
The case when the ratio of meaningless values is equal to 0 (i.e., \emph{sample-name}) 
represents the situation where the field was correctly filled for all the instances in the dataset.
 
Based on the ratio of meaningless values in Table~\ref{tab:dataset-info}, 
we find that the number of instances for
meaningless and valid values is imbalanced for most of the fields.
For example, the ratio of meaningless values for \emph{tissue} is 0.130.
The field age has more meaningless values with a ratio of 0.543.
The reason for this relatively high ratio could be that the completeness requirement (i.e., ``Required'')
of this field does not conform with the actual need in the real world;
that is, the field \emph{age} is not required when the actual concept
of ``age'' does not apply to a certain type of biomaterial (e.g., for  protein TM-1~\cite{song2017structural}).

The PEIS dataset contains the data filled through the web-based 
data entry form during the process of creating a new customer account.
The dataset was extracted from the database of our industrial partner. 
Similar to the NCBI dataset, each row in the table represents an instance and 
each column represents a form field.
We identified the mapping between column names in the table and field names in the data entry form 
using the available software documentation.

As shown in Table~\ref{tab:dataset-info}, the PEIS
dataset has 33 fields, 19 of which are required (including conditionally required).
In this dataset, nine of the required fields do not have missing/meaningless values 
(i.e., the ratio of meaningless values is 0). 
For the rest of the fields, the ratio of instances with missing or meaningless values  ranges from \numrange{0.113}{0.974}.
The reason behind having a high ratio of meaningless values in some fields, 
is that those fields are conditionally required.
They are rarely to be required in real scenarios, which leads to many missing values.

\subsubsection*{Dataset Preparation}
For the two datasets,
we consider all the required fields as targets since we aim to learn the conditions 
to relax them as optional (for avoiding meaningless values and
improving the overall data quality). 
However, we do not consider fields where the ratio of missing and meaningless values is 0, 
as they have no relaxation conditions to learn.
We split the dataset into three subsets containing 80\%, 10\%, and 10\% of
input instances based on their submission time, used respectively for training, tuning, and testing.  
The input instances (excluding submission time)
in the training set are used to train \approach.  
The validation set is used to decide the endorser threshold for each field 
following the strategy explained in section~\ref{sec:tuning}. 

As for the testing input instances, since
there is no information on the actual form filling order, 
we simulated two form filling orders for data entry, 
including ``sequential filling'' and ``partial random filling''.

The former corresponds to filling data entry forms in the default order, as
determined by the form \emph{tab sequence}, e.g., the navigation order
determined by the HTML attribute \texttt{tabindex} in web UI
designs~\cite{fowler2004web}.
It simulates the logical order many users follow to fill out forms, 
especially when they use a keyboard to navigate form fields~\cite{change2013microsoft}.
The latter represents the case when designers group some semantics-related fields together and add
controls to force users filling a group of fields sequentially~\cite{chen2011usher};
outside these groups, users can fill fields randomly. 

We simulated partial random filling by randomly
generating a field order for each testing input instance while
respecting the sequential order of the fields in the same group.
In the case where there is no grouping or controls in the form, 
the partial random filling scenario turns into to a (complete) random filling scenario.
The reason to simulate the partial random filling scenario is that 
by capturing the fields' grouping information,
this scenario is more realistic 
compared to a (complete) random filling scenario.

In both form
filling scenarios, the filled fields considered by \approach are the
fields that precede each target.
For each target field, we labeled as ``Optional'' the instances 
in which the target field contains missing or meaningless values; 
otherwise they are labeled as ``Required''.
``Optional'' and ``Required'' are the two classes 
that we consider as ground truth.

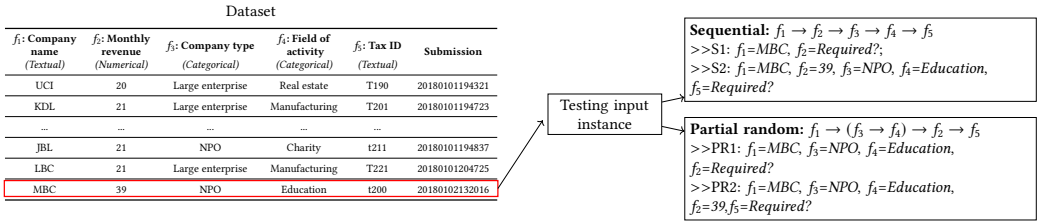
\begin{figure}[tb]
	\centering
	
\newcommand{\tikzmark}[2]{
	\tikz[overlay,remember picture,baseline] 
	\node [anchor=base] (#1) {$#2$};}

\newcommand{\DrawVLine}[3][]{\begin{tikzpicture}[overlay,remember picture]
	\draw[shorten <=0.3ex, #1] (#2.north) -- (#3.south);
	\end{tikzpicture}
}

\newcommand{\DrawCircle}[1][]{\begin{tikzpicture}[overlay,remember picture]
	\draw[black,fill=black] (0,0.5ex) circle (0.5ex);
	\end{tikzpicture}
}

\newcommand{\DrawRec}[1][]{\begin{tikzpicture}[overlay,remember picture]
	\draw[black,fill=black] (-0.5ex,0) rectangle ++(1ex, 1ex);
	\end{tikzpicture}
}

\newcommand{\DrawTriangle}[1][]{\begin{tikzpicture}[overlay,remember picture]
	\filldraw[fill=black] (90:1.2ex) -- (170:0.7ex) -- (10:0.7ex) --cycle;
	\end{tikzpicture}
}

\tikzstyle{round_rect} = [rectangle, rounded corners, minimum width=1.5cm, minimum height=0.5 cm,text centered, draw=black, fill=white!30]

\tikzstyle{rect} = [rectangle, minimum width=1.5cm, minimum height=0.8cm, text centered, text width=1.5cm, draw=black, fill=white!30]

\tikzstyle{text_node} = [rectangle, minimum width=0.5cm, minimum height=0.35cm, text centered, text width=4cm, draw=white, fill= white!30]

\tikzstyle{list_box} = [rect,align=left ,minimum width=3.19cm,text width=3.19cm, minimum height= 0.5cm]

\resizebox{1\textwidth}{!}{
\begin{tikzpicture}

\node (tab1) [shape=rectangle,draw=white] {
	\scriptsize
	\begin{tabular}{cccccc}
	\toprule
	
	\textbf{$f_1$: Company}&\textbf{$f_2$: Monthly}&\multirow{2}{*}{\textbf{$f_3$: Company type}}&\textbf{$f_4$: Field of }&\multirow{2}{*}{\textbf{$f_5$: Tax ID}}&\multirow{3}{*}{\textbf{Submission}}\\
	\textbf{name}&\textbf{revenue}&&\textbf{ activity}&&\\
	\emph{(Textual)}&\emph{(Numerical)}&\emph{(Categorical)}&\emph{(Categorical)}&\emph{(Textual)}\\
	\midrule
	
	UCI &20&Large enterprise&Real estate&T190&20180101194321 \\
	\midrule
	KDL&21&  Large enterprise&Manufacturing&T201 &20180101194723 \\ 
	\midrule
	...&...&  ...&...& ...  \\ 
	\midrule
	JBL&21&  NPO&Charity&t211 &20180101194837 \\ 
	\midrule
	LBC&21& Large enterprise&Manufacturing&T221 &20180101204725 \\ 
	\midrule
	MBC &39&NPO&Education &t200&20180102132016 \\

	\bottomrule
\end{tabular}
};

\node (dataset) [ text_node, above=of tab1, yshift=-1cm] {Dataset};
\node (test-ed) [ text_node, draw=red, thick, text width=2.3cm, below= of tab1,minimum width=11cm, minimum height=0.35cm,yshift=1.58cm,fill opacity=0 ] { };
\node (test) [ text_node, draw=black, text width=2.3cm, right= of tab1] { Testing input instance };

\draw [arrow,->] (test-ed) ++(5.5, 0)--++(1, 1.5) (test);

\node (seq) [ text_node, align= left, draw=black, text width=7.6cm, right= of test, xshift=-0.5cm, yshift= 1.2cm] { \textbf{Sequential:} $f_1\rightarrow f_2 \rightarrow f_3 \rightarrow f_4 \rightarrow f_5$\\
$>>$S1: $f_1$=\emph{MBC}, $f_2$=\emph{Required?};\\
$>>$S2: $f_1$=\emph{MBC}, $f_2$=\emph{39}, $f_3$=\emph{NPO}, $f_4$=\emph{Education}, $f_5$=\emph{Required?}};

\node (rand) [ text_node, align= left, draw=black, text width=7.6cm, right= of test, xshift=-0.5cm, yshift= -1.2cm] { \textbf{Partial random:} $f_1\rightarrow (f_3 \rightarrow f_4 )\rightarrow f_2 \rightarrow f_5$\\
	$>>$PR1: $f_1$=\emph{MBC}, $f_3$=\emph{NPO}, $f_4$=\emph{Education}, $f_2$=\emph{Required?}\\
	$>>$PR2: $f_1$=\emph{MBC}, $f_3$=\emph{NPO}, $f_4$=\emph{Education}, $f_2$=\emph{39},$f_5$=\emph{Required?}};

\draw [arrow,->] (test) -- (seq);
\draw [arrow,->] (test) -- (rand);

\end{tikzpicture}
}

 	\caption{Example of Filling Orders}
	\label{fig:eval}
\end{figure}

\subsubsection*{Dataset Preparation - Application Example}
\figurename~\ref{fig:eval} illustrates an example of application of our dataset preparation method. 
The table on the left-hand side of the picture represents the information submitted during the data entry session 
through the data entry form introduced in our motivating example in section~\ref{sec:definition}.
We split this dataset into a training set (80\% of instances), a
tuning set (10\% of instances), and a testing set (10\% of
instances); 
let us assume the last row in the table is an instance in the testing set.
The testing set is then processed to simulate the two form filling
scenarios.  
The sequential filling scenario uses the filling order following
the \texttt{tabindex} value of the form fields.
Assuming the \texttt{tabindex} order for the example is
$f_1\rightarrow f_2 \rightarrow f_3 \rightarrow f_4 \rightarrow f_5$, we can generate 
two test instances S1 and S2 (shown in the top right box of \figurename~\ref{fig:eval}) to predict the completeness requirement of $f_2$ and $f_5$, respectively.
As for the partial random filling scenario, this scenario takes into account 
the controls or grouping of fields specified by the designer.
For example, let us assume that ``$f_3$ : company type '' and ``$f_4$:
field of activity'' belong to the same group of fields named
``Business activities'': this means that $f_3$ and $f_4$ should be
filled sequentially. A possible filling order, randomly generated
taking into account this constraint is then  $f_1 \rightarrow (f_3
\rightarrow f_4) \rightarrow f_2 \rightarrow f_5 $. The bottom right
box in the figure shows the corresponding generated test instances PR1 and PR2.

\subsubsection*{Implementation and Settings}
We implemented \approach as a Python program.
We performed the experiments on the NCBI dataset with a computer running macOS 10.15.5 with a \SI{2.30}{\giga\hertz}
Intel Core i9 processor with \SI{32}{\giga\byte} memory. As for the
experiments on the PEIS dataset\footnote{Due to the data protection
	policy of our partner, we were obliged to run the experiments on the
	PEIS dataset using an on-premise, dedicated server that, however,
	could not be used to store external data (like the NCBI dataset).}, we performed them on a server running CentOS 7.8 on a \SI{2.60}{\giga\hertz} Intel Xeon E5-2690 processor with \SI{125}{\giga\byte} memory.

\subsection{Effectiveness (RQ1)}
\label{sec:rq1}

To answer RQ1, we assessed the accuracy of \approach in predicting the correct completeness requirement for each target field in the
dataset.
To the best of our knowledge, there are no implementations of
techniques for automatically relaxing completeness requirements;
therefore, we selected as baselines two rule-based algorithms that
can be used to solve the form filling completeness requirements
relaxation problem: 
 \emph{association rule mining} (ARM)~\cite{martinez2019using} and
  \emph{repeated incremental pruning to produce error reduction}
  (Ripper); these rule-based algorithms can provide form filling relaxation suggestions under different filling orders.
ARM mines association rules having the format ``if \emph{antecedent} then \emph{consequent}'' with a minimal level of support and confidence,
where the antecedent includes the values of certain fields and the
consequent shows the completeness requirement of a target field for a given antecedent. 
ARM matches the filled fields with the antecedents of mined association rules,
and suggests the consequent of the matched rules.
Ripper is a propositional rule-based classification algorithm~\cite{cohen1995fast}; it creates
a rule set by progressively adding rules to an empty set until all the positive instances are covered~\cite{linares2014using}. 
Ripper includes also a pruning phase to remove rules 
leading to bad classification performance.
Ripper has been used 
in a variety of classification tasks in software engineering~\cite{song2018comprehensive,ghotra2015revisiting}.
Similar to ARM, Ripper suggests the consequent of the matched rule to users.
 
\paragraph{Methodology}
We used Precision ($\mathit{Prec}$), Recall ($\mathit{Rec}$), Negative Predictive Value ($\mathit{NPV}$), and Specificity ($\mathit{Spec}$) to assess the accuracy of different algorithms.
These metrics can be computed from a confusion matrix that classifies the prediction results into true positive (TP), false positive (FP), true negative (TN), and false negative (FN). In our context, TP means that a field is correctly predicted as required,
FP means that a field is 
misclassified as required, 
TN means that a field is correctly predicted as optional,
and FN
means that a field is misclassified as optional.
Based on the confusion matrix, we have $\mathit{Prec}=\frac{\mathit{TP}}{\mathit{TP}+\mathit{FP}}$, $\mathit{Rec}=\frac{\mathit{TP}}{\mathit{TP}+\mathit{FN}}$,
$\mathit{NPV}=\frac{\mathit{TN}}{\mathit{TN}+\mathit{FN}}$, and
$\mathit{Spec}=\frac{\mathit{TN}}{\mathit{TN}+\mathit{FP}}$.
Precision is the ratio of correctly predicted required fields over all the fields predicted as required.
Recall is the ratio of correctly predicted required fields 
over the number of actual required fields.
NPV represents the ratio of correctly predicted optional fields
over all the fields predicted as optional.
Finally, specificity represents the ratio of correctly predicted optional fields over the number of actual optional fields.
 
We chose these metrics because they can evaluate the ability of an algorithm in predicting both required fields (using precision and recall) and optional fields (using NPV and specificity).
A high value of precision and recall means that an algorithm can correctly predict most of required fields (i.e., the positive class); hence, we can avoid business loss caused by missing information.
A high value of NPV and specificity means that an algorithm can correctly predict most of the optional fields (i.e., the negative class);
users will have fewer unnecessary constraints during form filling. In other words, we can avoid users filling meaningless values which may affect the data quality. 

In our application scenario, we aim to successfully  relax a set of obsolete required fields to ``optional'', 
while keeping the real required fields.
Therefore, \approach needs to get high precision and recall values, which can preserve most of real required fields to avoid business loss.
Meanwhile, the NPV value should be high, which means \approach can correctly avoid users filling meaningless values by relaxing the completeness requirements.
Concerning the specificity, a relatively low value is still useful.
For instances, a specificity value of 50\% means \approach can reduce by half the data quality issues caused by meaningless values.
 
In the case of ARM, we set the minimum acceptable support and
confidence to 5 and 0.3, respectively, as done in previous work~\cite{martinez2019using,belgacem2022machine} in
which it was applied in the context of form filling.

\subsubsection*{Results}

 \begin{table*}[tb]
 	\caption{Effectiveness for Form Filling
 		Relaxation}
 	\centering
 	\scriptsize
 	\pgfplotstabletypeset[
every head row/.style={
	before row={\toprule
		\multirow{2.5}{*}{}&\multirow{2.5}{*}{Alg.}&\multicolumn{4}{c}{Sequential}&\multicolumn{4}{c}{Partial Random}&\multirow{1}{*}{Train}&\multicolumn{2}{c}{Predict (\si{\milli\s})}\\
		\cmidrule(r){3-6}
		\cmidrule(l){7-10}
	},
	after row=\midrule,
},
every last row/.style={
	after row=\bottomrule
},
every nth row={3}{before row=\midrule},
columns/Dataset/.style ={column name=}, 
columns/Algorithm/.style ={column name=}, 
columns/Prec-S1/.style={column name=Prec}, 
columns/Rec-S1/.style ={column name=Rec},
columns/NPV-S1/.style ={column name=NPV},
columns/Spec-S1/.style ={column name=Spec},
columns/Prec-R1/.style={column name=Prec}, 
columns/Rec-R1/.style ={column name=Rec},
columns/NPV-R1/.style ={column name=NPV},
columns/Spec-R1/.style ={column name=Spec},
columns/Train/.style ={column name=(\si{\s})},
columns/P-Avg/.style ={column name=avg},
columns/P-Range/.style ={column name=min--max},
col sep=comma,
string type,
]{data/rq-applicability.csv}

  	\label{tab:rq1}
 \end{table*}

Table~\ref{tab:rq1} shows the accuracy of the various algorithms
for the two form filling scenarios.
\approach substantially outperforms Ripper in terms of  precision and
recall scores (i.e., columns $\mathit{Prec}$ and $\mathit{Rec}$) 
 for both sequential filling and partial random filling scenarios in
 both datasets (ranging from \SIrange{+13}{+32}{\pp} in terms of precision score and
 from \SIrange{+15}{+35}{\pp} in terms of recall score). 
When we compare \approach with ARM, they have similar results in terms
of precision and recall scores on the NCBI dataset;
however, \approach performs much better than 
ARM on the PEIS dataset (by at least \SI{+16}{\pp}  in terms of
precision score and \SI{+17}{\pp} in terms of 
recall score).

When looking at the $\mathit{NPV}$ and specificity scores, on the NCBI
dataset 
\approach and Ripper have the same specificity value for sequential filling;
however, \approach can provide more accurate suggestions since it outperforms Ripper in terms
of NPV score with an improvement of \SI{+74}{\pp}. 
Concerning the partial random filling scenario on the NCBI
dataset, \approach outperforms Ripper
by \SI{+51}{\pp} and \SI{+21}{\pp} in terms of NPV and specificity scores, respectively.
On the same dataset, when comparing \approach with ARM, the results shows that \approach
always outperforms ARM for both form filling scenarios from
\SIrange{+10}{+37}{\pp} in terms of NPV score 
and from \SIrange{+4}{+9}{\pp} in terms of specificity score.
As for the PEIS dataset, 
for sequential filling \approach substantially outperforms the two
baselines from \SIrange{+12}{48}{\pp} in terms of NPV score and 
from \SIrange{+33}{+38}{\pp} in terms of specificity score.
For partial random filling, Ripper achieves the highest NPV score,
outperforming \approach by \SI{+9}{\pp}; however,
\approach  outperforms both baselines in terms of specificity score by
\SIrange{+8}{+39}{\pp}.

Looking at the specificity score when applying \approach on PEIS and NCBI datasets,
we can notice a difference ranging from \SIrange{+27}{+42}{\pp}.
 This difference means that \approach can find more optional values in the
PEIS dataset than in the NCBI dataset.  We believe the main reason
behind this difference is the quality of the training set.  We recall
PEIS is a proprietary dataset from the banking domain. Data entry
operators in the bank follow corporate guidelines for recommended
values to be used when a field is not applicable, e.g., special
characters like `@' or `\$' (see section~\ref{sec:pre-processing}),
resulting in higher quality data than the NCBI dataset. The latter, in
fact, is a public dataset where anyone can submit data using the
corresponding data entry form.  Users do not follow any rule to insert
special values when a field is not applicable.  For this reason,
the endorser module of \approach tends to remove more likely
inaccurate suggestions, predicting only optional fields with high
confidence.  This explains the high value of NPV in the NCBI dataset,
which is \SI{+19}{\pp} higher than that in the PEIS dataset for the
sequential filling scenario and \SI{+1}{\pp} higher for the random
filling scenario.

We applied Fisher’s exact test with a level of significance $\alpha = 0.05$ to assess the statistical 
significance of differences between \approach and the baselines.
The null hypothesis is that there is no significant difference between the prediction results of \approach and a baseline algorithm on the test instances. 
Given the output of each algorithm on the test instances we used
during our evaluation, 
We created contingency tables summarising the decisions of \approach vs ARM and \approach vs Ripper for each form-filling scenario.
Each contingency table represents the relationship between \approach and the other baseline in terms of frequency counts of the possible outputs (0: ``Optional'' and 1: ``Required''). 
In other words, the contingency table counts the number of times both algorithms provide the same prediction (i.e., both predict a test instance as 0 or 1), and the number of times they have different prediction outputs (i.e., one algorithm predicts as 1 but the other predicts 0, and vice versa).
These contingency tables are then used by Fisher’s exact test to compute the p-value in order to reject or accept the null hypothesis.
The result of the statistical test shows that \approach always achieves a significantly higher number of correct predictions than the baselines
for the two form-filling scenarios on both datasets ($\textit{p-value}<0.05$).

These results have to be interpreted with respect to the usage scenario of a form filling 
relaxation tool.
Incorrect suggestions can affect the use of data entry forms
and the quality of input data. 
The $\mathit{NPV}$ and specificity values achieved 
by \approach show that its suggestions can help users accurately relax the completeness
requirement by \SIrange{20}{64}{\%} of the fields in data entry forms. Meanwhile, \approach can correctly preserve most ($\ge 97\%$) of the required fields required to be filled to avoid
missing information (as indicated by the high precision and
recall scores).
 
\emph{The answer to RQ1 is that \approach performs significantly better than the baseline algorithms. \approach can correctly relax at least 20\% of required fields
(with an NPV value above {\LACQUERSeqNPVProp}), while preserving the
completeness constraints on most of the truly required fields (with a recall value over {\LACQUERRandRec} and precision over {\LACQUERSeqPrec}).}

 \subsection{Performance (RQ2)}
\label{sec:rq:performance}

To answer RQ2, we measured the time needed to perform the training and predict 
the completeness requirement of target fields. 
The training time evaluates the ability of \approach to efficiently update its models when new input instances are added daily to the set of historical input instances.
The prediction time evaluates the ability of \approach to timely suggests the completeness requirement during the data entry phase. 

\subsubsection*{Methodology}

We used the same baselines and form-filling scenarios used for RQ1. 
The training time represents the time needed to build 
BN models (for \approach) or learn association rules (for ARM and Ripper). 
The prediction time is the average time needed to provide suggestions
for target fields.
We deployed \approach and baselines locally to avoid the impact of the data transmission time when assessing the prediction time. 

\subsubsection*{Results}
The results are presented in columns \emph{Train} and \emph{Predict} in Table~\ref{tab:rq1}. 
 Column \emph{Train} represents the training time in seconds.
Column \emph{Predict} contains two sub-columns representing the
average time and the minimum/maximum time (in milliseconds) needed
to make a prediction on one test instance.

As shown in Table~\ref{tab:rq1},
Ripper has the highest training time for the NCBI dataset with \SI{\RipperNCBITrain}{\s}. 
The training time of \approach (\SI{\LACQUERNCBITrain}{\s}) is between that of Ripper (\SI{\RipperNCBITrain}{\s}) and ARM (\SI{\ArmNCBITrain}{\s})
on the NCBI dataset.
For the PEIS dataset, the training time of Ripper and ARM is equal to 
\SI{\RIPPERPROPTrain}{\s} and \SI{\ARMPROPTrain}{\s}, respectively;
the training time of 
\approach is the highest:
\SI{\LACQUERPROPTrain}{\s} (less than 20 minutes).
 
In terms of prediction time, \approach takes longer than ARM and Ripper to predict 
the completeness requirement of a field.  On average, \approach takes
\SI{\LACQUERNCBIAvg}{\milli\s} and \SI{\LACQUERPROPAvg}{\milli\s} on
the NCBI and PEIS datasets, respectively.
The prediction time of ARM and Ripper depends on the number of rules used for matching the filled fields:
the smaller the number of
rules  the shorter the prediction time.
For \approach, the prediction time depends mostly on the complexity of BNs used when predicting.
Such complexity can be defined in terms of the number of nodes and the
number of dependencies among the different nodes in the BNs.  
 
Taking into account the usage of our approach, the results can be
interpreted as follows.  Since the training phase occurs
\emph{offline} and \emph{periodically} to train different BN models,
the training time of \SI{\LACQUERPROPTrain}{\s} is acceptable from a
practical standpoint; it allows for the daily (or even hourly) training of \approach in contexts (like enterprise software) with thousands of entries every day.
Since \approach needs to be used during data
entry, a short prediction time is important to preserve the
interactive nature of a form-filling relaxation tool. 
The prediction time of \approach is acceptable according to
human-computer interaction principles~\cite{heeter2000interactivity},
which prescribe a response time lower than \SI{1}{\s} for tools that
provide users with a seamless interaction.  In addition, this
prediction time is also comparable to the one achieved by our previous
work on automated form filling ~\cite{belgacem2022machine}.  Hence,
\approach can be suitable for deploying in real enterprise systems.

\emph{The answer to RQ2 is that the performance of \approach, with a
	training time per form below 20 minutes  and a prediction time of at
	most \SI{839}{\milli\s} per target field, is suitable for practical
	application in data-entry scenarios.}

 \subsection{Impact of SMOTE and Endorser (RQ3) }
\label{sec:rq3}

\approach is based on two main modules:
\begin{inparaenum}[(1)]
	\item SMOTE oversampling module, which tries to solve the class imbalance problem by synthetically 
	creating new minor class instances in the training set (section~\ref{sec:model-building}), and 
	\item the endorsing  module, which implements a heuristic that aims to 
	keep only the optional predicted instances with a certain level of confidence. 
\end{inparaenum}
To answer this RQ we assessed the impact of these two modules on the 
effectiveness of \approach. 

\subsubsection*{Methodology}

We compared the effectiveness of \approach
with three variants representing all the possible 
configurations of \approach:  \approach-S, \approach-E, and \approach-SE. 
\approach-S represents the configuration where the SMOTE oversampling module is disabled
and \approach provides predictions based on the imbalanced training set. 
\approach-E denotes the configuration where the endorser module is disabled and \approach
directly returns the predictions to the user without checking whether the predictions have the required confidence in
predicting fields as optional. 
\approach-SE is the configuration where both modules are disabled; this variant corresponds to the case where 
we use a plain BN. 
The different configurations are shown in Table~\ref{tab:component}
under column \emph{Module},
where the two sub-columns \emph{S} and \emph{E} refer to 
the two modules ``\textbf{S}mote'' and ``\textbf{E}ndorser''. 
We used symbols `\ding{51}' and `\ding{55}' to specify whether a variant includes or not a certain module.
\approach was run in its vanilla
version as well as the additional variants using the same settings and evaluation metrics
as in RQ1.

\subsubsection*{Results}

\begin{table}[tb]
	\centering
		\scriptsize
	\caption{Effectiveness of \approach with Different Modules}
	\label{tab:component}
	\resizebox{\columnwidth}{!}{\begin{filecontents*}{data/rq-component.csv}
ID, Local, Filter, PRE-S1,REC-S1,NPV-S1,SPE-S1,PRE-R1,REC-R1,NPV-R1,SPE-R1, PRE-S2,REC-S2,NPV-S2,SPE-S2,PRE-R2,REC-R2,NPV-R2,SPE-R2	
\approach-SE, \ding{55}, \ding{55}, 0.78, 0.88, 0.36,0.32 ,0.84 ,0.89 ,0.44 ,0.33 ,0.90 ,0.92 ,0.70 ,0.67 , 0.96,0.96 ,0.75 ,0.58 
\approach-E,  \ding{51}, \ding{55}, 0.78, 0.82, 0.56, 0.32, 0.85,0.77 ,0.39 ,0.65 ,0.89 ,0.94 ,0.67 ,0.76 ,0.92 ,0.88 ,0.57 , 0.85
\approach-S,  \ding{55}, \ding{51}, 0.76,0.98 ,0.51 ,0.19 ,0.83 ,0.98 ,0.66 ,0.22 , 0.88,0.99 ,0.70 ,0.55 ,0.91 ,0.99 , 0.77, 0.52
\approach,    \ding{51}, \ding{51}, 0.76, 0.98, 0.91, 0.20, 0.84, 0.98,0.76 ,0.37 , 0.89, 0.99, 0.74, 0.64, 0.90, 0.97, 0.75, 0.64
\end{filecontents*}
\pgfplotstabletypeset[
	every head row/.style={
		before row={\toprule
			\multirow{4.5}{*}{ID}&\multicolumn{2}{c}{\multirow {2.5}{*}{Module}}&\multicolumn{8}{c}{NCBI}&\multicolumn{8}{c}{PEIS}\\
			\cmidrule(lr){4-11}
			\cmidrule(lr){12-19}
			&&&\multicolumn{4}{c}{Sequential}&\multicolumn{4}{c}{Partial
                          Random}&\multicolumn{4}{c}{Sequential}&\multicolumn{4}{c}{Partial
                          Random}\\
			\cmidrule(lr){2-3}
			\cmidrule(lr){4-7}
			\cmidrule(lr){8-11}
			\cmidrule(lr){12-15}
			\cmidrule(l){16-19}
		},
		after row=\midrule,
	},
	every last row/.style={
		after row=\bottomrule
	},
	columns/ID/.style = {column name=},
	columns/Local/.style = {column name=S},
	columns/Filter/.style = {column name=E},
	columns/PRE-S1/.style ={column name=Prec}, 
	columns/REC-S1/.style={column name=Recall},  
	columns/NPV-S1/.style={column name=NPV},  
	columns/SPE-S1/.style={column name=Spec},  
	columns/PRE-R1/.style ={column name=Prec}, 
	columns/REC-R1/.style={column name=Recall},  
	columns/NPV-R1/.style={column name=NPV},  
	columns/SPE-R1/.style={column name=Spec},   
	columns/PRE-S2/.style ={column name=Prec}, 
	columns/REC-S2/.style={column name=Recall},  
	columns/NPV-S2/.style={column name=NPV},  
	columns/SPE-S2/.style={column name=Spec},  
	columns/PRE-R2/.style ={column name=Prec}, 
	columns/REC-R2/.style={column name=Recall},  
	columns/NPV-R2/.style={column name=NPV},  
	columns/SPE-R2 /.style={column name=Spec},  
	col sep=comma,
	string type,
	]{data/rq-component.csv}
}

 \end{table}

As shown in Table~\ref{tab:component}, 
both modules have an impact on the effectiveness of \approach.
The SMOTE oversampling module improves the ability of BNs to identify 
more optional fields;
it  improves
the specificity score of a plain BN by at least \SI{+9}{\pp} on the two datasets (\approach-E vs \approach-SE),
except for the sequential filling scenario in the NCBI dataset
where the specificity score stays the same.
  The endorser module mainly removes inaccurate optional predictions  
 and keeps them as required to prevent missing information. 
 This module leads to an increase in the recall value compared to the plain BN (\approach-SE vs \approach-S); it increases by at least  \SI{+9}{\pp} for the NCBI dataset in both scenarios. 
 The improvement is smaller for the PEIS dataset where the recall increases by
  \SI{+7}{\pp} and \SI{+3}{\pp} for sequential and random filling scenarios, respectively. 
 The endorser module affects also specificity, which decreases 
 by at most \SI{13}{\pp} for both datasets when the endorser is used. 
 The reason behind such decrease is that the endorser module removes possibly inaccurate predictions.

Comparing the results of \approach (with both modules  enabled) with a plain BN
(i.e, \approach-SE) on the NCBI dataset, the former improves NPV by \SI{+55}{\pp} (0.91 vs 0.36)
for the sequential filling scenario and by \SI{+32}{\pp} (0.76 vs 0.44) for the random filling scenario. 
Since the endorser module considers the non-endorsed instances as required, it also increases 
recall by\SI{+10}{\pp} and \SI{+9}{\pp} for sequential and random filling scenarios, respectively. 
For the PEIS dataset, we find a slight increase in NPV of \SI{+4}{\pp} and an increase of \SI{+6}{\pp}
for recall with sequential filling. 
For the partial random filling scenario, we notice that both \approach and \approach-SE have similar
results, except for a higher specificity value \SI{+6}{\pp}
and a lower precision value of \SI{-6}{\pp} for \approach. 
This loss in precision is expected since \approach keeps the default
completeness requirement (i.e., required) for an instance 
for which the prediction confidence is low  
(i.e., the probability is lower than a threshold in endorser). 
These instances may include some truly optional cases with low confidence in the prediction; hence
considering them as optional may slightly reduce the precision value.

\emph{The answer to RQ3 is that the SMOTE oversampling module and the
endorser module improve the effectiveness of \approach.}

 \subsection{Threats to Validity}
\label{sec:threats}

To increase the generalizability of our results, 
\approach should be further evaluated on different datasets from different domains.
To partially mitigate this threat, we evaluated \approach on two datasets
with different data quality: 
the PEIS dataset, which  is proprietary and of high quality, and 
the NCBI dataset, which is public and was obtained from an environment
with looser data quality controls. 

The size of the pool of training sets is a common threat to all AI-based approaches.
We do not expect this problem to be a strong limitation of \approach since it
targets mainly enterprise software systems that can have thousands of entries per day.

Since  \approach needs to be run online during the data entry session, 
it  is important to ensure seamless interaction with users. 
In our experiments (section~\ref{sec:rq:performance}),
\approach  was deployed locally. The response time of its prediction complies with human-computer interaction 
standards.
However, the prediction time depends on the deployment method (e.g., local deployment or cloud-based).
This is not necessarily a problem since different engineering methods can help reduce 
prediction time such as parallel computing and a cache for storing previous predictions.

\subsection{Data Availability}\label{sec:data-availability}

The implementation of \approach,
the  NCBI dataset, and the scripts used for the evaluation are available at \url{https://figshare.com/articles/software/LACQUER-replication-package/21731603}; 
\approach is distributed under the MIT license. 
The PEIS dataset cannot be distributed due to an NDA.

\section{Related work}
\label{sec:related_work}
In this section, we discuss the work related to our approach. 
First, we review the existing approaches dealing with adaptive forms.
Next, we provide a detailed comparison between \approach and LAFF. We
conclude the section by presenting some tangential works that use BN to solve software engineering problems.

\subsection{Adaptive Forms}
\label{sec:rw-adaptive-forms}

The approach proposed in this paper is mainly related to approaches
that implement adaptive forms for producing context-sensitive
form-based interfaces.  These approaches progressively add (remove)
fields to (from) the forms depending on the values that the user
enters.  They use \emph{form specification
	languages}~\cite{frank1998adaptive} or \emph{form definition
	languages}~\cite{bohoj2011adapforms} to allow form designers to
describe the dynamically changing behaviour of form fields. Such a
behavior is then implemented through dedicated graphical user
interface programming languages (such as
Tcl/Tk)~\cite{thistlewaite1996active} or through server-side
validation~\cite{bohoj2011adapforms}.  The dynamic behaviour of a form
has also been modeled using a declarative, business process-like
notation (DCR - Dynamic Condition Response
graph~\cite{stromsted2018dynamic}), where nodes in the graph represent
fields and edges show the dynamic relations among fields (e.g.,
guarded transitions); the process declarative description is then
executed by a process execution engine that displays the form. 
However, all
these works assume that designers already have a \emph{complete and
	final} set of completeness requirements describing the adaptive
behaviour of the form during the design phase, which can be expressed
through (adaptive) form specification/definition languages or tools.
In contrast, \approach can automatically learn the different
completeness requirements from the historical input instances filled
by users, without requiring any knowledge from the form designers.

Although some approaches~\cite{elbibas2004developing,albhbah2010using}
try to automatically generate data entry forms based on the schema of
the database tables linked to a form (e.g., using column name and primary
keys), they can only generate some ``static'' rules for fields.  For
example, if a column is ``not null'' in the schema, they can set the
corresponding field in the form as (always) required.  In contrast,
\approach aims to learn conditions from the data so that completeness
requirements of form fields can be \emph{automatically and
	dynamically} relaxed during new data entry sessions.

\subsection{Comparing \approach with LAFF}
\label{sec:revc-appr-with}

The overall architecture (including the use of the endorser module) of
\approach has been inspired by LAFF, a recent approach for
automated form filling of data entry forms~\cite{belgacem2022machine}.
In this subsection, we explain the similarities and differences between the two approaches.
\subsubsection*{Similarities between LAFF and \approach}
Both LAFF and \approach are approaches that can be used during the
form filling process.
The main similarities between these approaches derive from the
main challenges of form filling, i.e.,  dealing
with (1) an arbitrary filling order and (2)  partially filled forms.

The first challenge arises from the fact that users can fill a data
entry form following an arbitrary order.  Therefore, the filled fields
(i.e., the features in our ML models) and the target field keep changing, leading to a
large number of feature-target combinations.  To avoid training a
separate machine learning model on each feature-target combination, in
this work, we are inspired by LAFF and use BNs to mine the
relationships between filled fields and the target field.

As for the second challenge, LAFF addresses it using an endorser
module. The main idea of the endorser module is to avoid providing
inaccurate suggestions to the user when the form does not contain
enough information for the model.  Avoiding inaccurate
suggestions is important for both approaches to gain the trust of
users;  for example, wrongly determining to relax a required field by
making it optional may lead to missing information, thus hindering 
data completeness.  For this reason, the second similarity between
LAFF and \approach is the use of an endorser module.

\subsubsection*{Differences between LAFF and \approach}
\label{sec:differences}

	\tablename~\ref{tab:diff-table} shows the main differences between \approach and LAFF in terms of \textit{goal}, \textit{challenges}, 
	\textit{preprocessing}, \textit{model building}, and \textit{prediction}.

 The main \emph{goal} of \approach is to determine the completeness requirements
	of form fields. 
	In contrast,  LAFF provides form-filling suggestions for the values to be filled in categorical fields.
Concerning the \emph{challenges}, in addition to the shared
        ones discussed above, the relaxing completeness requirement
        problem has its own challenge when the dataset is highly
        imbalanced. We addressed this challenge in \approach by applying SMOTE.

The \emph{preprocessing} step of the two approaches is completely different.
	Specifically, LAFF removes all textual fields from the data. 
	In contrast,  \approach transforms the values in textual
        fields into binary values. After the preprocessing, 
	textual fields can only have one of two values: ``Required'' and
        ``Optional''.
        Moreover, the preprocessing step of \approach identifies
        meaningless values and replaces the matched values in the data
        with the value ``Optional'' (see section~\ref{sec:pre-processing}). 

As for the \emph{model building} phase, LAFF and \approach create a different set of BN models. 
	LAFF creates $k+1$ models, including a global model  
	and $k$ local models. 
	The global model represents the BN created on the whole
        training data; 
	the $k$ local models are the BNs created based on the clusters
        of training data that share similar characteristics.
    The optimal number of clusters $k$ is automatically determined with the elbow method. 
	\approach  creates $n$ models where $n$ represents the number of fields (targets) in the data entry form.

Finally, the differences regarding the \emph{prediction} phase
  can be viewed from two perspectives: the type of targets and the endorser module. 
	Concerning the target, LAFF only predicts possible values for categorical fields, no matter whether this field is optional or required. 
	In contrast, \approach targets all types of required fields (e.g., textual, numerical, and categorical fields) to relax their completeness requirements.
The endorser modules of LAFF and \approach differ as follows:
        \begin{itemize}
        \item The endorser module of LAFF endorses predictions based on two heuristics: the prediction confidence and the dependencies between the filled fields
	and the target. In contrast, the endorser of \approach is
        based only on the prediction confidence.
      \item LAFF uses a threshold to be determined manually, based on
        domain expertise, to endorse the prediction whereas  \approach includes a phase to automatically determine the threshold for each target.
        \end{itemize}

\begin{table}[tb]
  \caption{Main differences between LAFF and \approach}
\centering
	\scriptsize
	\begin{tabular}{p{1.25cm} p {0.5cm}p{5.25cm }p{5.25cm}}
	\toprule
	&&\multicolumn{1}{c} {LAFF} &\multicolumn{1}{c} {\approach}\\
	\midrule

	\multicolumn{2}{ l}{\multirow{2.5}{*}{Goal} } & 
	 \vspace{-0.5\baselineskip}
	\begin{itemize}
	\item Providing form-filling suggestions for the values to be filled in categorical fields
	\end{itemize}
	\nointerlineskip
	&
	\vspace{-0.5\baselineskip}
	\begin{itemize}
		\item Determining the completeness requirements of form
	\end{itemize}
    \nointerlineskip \\
	\cmidrule{1-4}
	
	\multicolumn{2}{ l}{\multirow{3}{*}{Challenge}} & 
	 \vspace{-0.5\baselineskip}
	\begin{itemize}
	  \item Arbitrary filling order
	\item Partial filling
	\end{itemize}
	\nointerlineskip
	& 
	\vspace{-0.5\baselineskip}
	\begin{itemize}
		\item Arbitrary filling order
		\item Partial filling 
		\item Highly imbalanced dataset
	\end{itemize}
	\nointerlineskip \\
	\cmidrule{1-4}
	
	\multicolumn{2}{ l}{\multirow{3}{*}{Preprocessing}} &  
	\vspace{-0.5\baselineskip}
	\begin{itemize}
	 \item  Textual fields are removed
	\end{itemize}
	\nointerlineskip 
	&
	\vspace{-0.5\baselineskip}
	\begin{itemize}
	\item Values in textual fields are transformed into binary values (``Required'' or ``Optional'')
	\item Meaningless values are identified and replaced with the value ``Optional''
	
	\end{itemize}
\nointerlineskip  \\
	\cmidrule{1-4}
	
	\multicolumn{2}{ l}{\multirow{4}{*}{Model building}} &  
	\vspace{-0.5\baselineskip}
	\begin{itemize}
	  \item Creates $k+1$ models including a global model and $k$ local models (one model for each cluster of data)

	 	\end{itemize}
 \nointerlineskip  
	& 
	\vspace{-0.5\baselineskip}
	\begin{itemize}
	\item  Creates $n$ models, one model for each field (target)
\end{itemize}
  \nointerlineskip   \\
	\cmidrule{1-4}
	\multirow{8}{*}{Prediction} & \multirow{3}{*}{Target} &
	\vspace{-0.5\baselineskip}
	\begin{itemize}
		 \item Categorical field
		 \item LAFF can predict the value for both optional and required fields
	\end{itemize}
	 \nointerlineskip  
	  & 
	  \vspace{-0.5\baselineskip}
	  \begin{itemize}
         \item All textual, numerical, and categorical fields can be targets
		 \item Required field
		\end{itemize}
	 \nointerlineskip  
	  \\ \cline{2-4}
	
	& \multirow{4}{*}{Endorser} &
	 \vspace{-0.5\baselineskip}
	\begin{itemize}
	     \item Use two heuristics based on prediction confidence and dependencies between filled fields and the target
		 \item The value of the threshold is manually decided based on domain expertise
		 	\end{itemize}
	 \nointerlineskip & 
	 \vspace{-0.5\baselineskip}
	 \begin{itemize}
		\item The endorser is based only on the prediction confidence
		\item The value of the threshold is automatically determined during the threshold determination
	\end{itemize}
\nointerlineskip \\

	\bottomrule
\end{tabular}

 	\label{tab:diff-table}
\end{table}

\subsection{Using Bayesian Networks in Software Engineering Problems}
\label{sec:UseBNinSE}
Besides LAFF, BNs have been applied to different software
  engineering problems spanning over a wide range of software
  development phases, such as project management (e.g., to estimate
  the overall contribution that each new software feature to be
  implemented would bring to the company~\cite{mendes2018using}),
  requirement engineering (e.g., to predict the requirement complexity
  in order to assess the effort needed to develop and test a
  requirement~\cite{sadia2022bayesian}), implementation (for code
  auto-completion \cite{proksch2015intelligent}), quality assurance
  (e.g., for defect
  prediction~\cite{jeet2011bayesian,dejaeger2012toward}), and software
  maintenance~\cite{rey2023bayesian}. 

The main reason to use BN in software engineering (SE) problems is the ability of BN to address the 
challenges of dealing with ``large volume datasets'' and ``incomplete data entries''.
First, software systems usually generate large amounts of data~\cite{rey2023bayesian}.
For instance, to improve software maintenance, companies need to
analyze large amounts of software execution data (e.g., traces and logs) 
to identify unexpected behaviors such as performance degradation.
To address this challenge, \citet{rey2023bayesian} used BN to build an analysis model on the data, since
BN can deal with large datasets and high-dimensional data while
keeping the model size small and the training time low.  
Second, incomplete data is a common problem in SE~\cite{del2016bayesian, okutan2014software}.
For example, some metrics in defect prediction datasets might be missing for some software modules.
To solve this challenge, \citet{okutan2014software} and \citet{del2016bayesian} used BN to train prediction models, because of 
its ability to perform inference with incomplete data entries.
These two challenges confirm our choice of using BN to solve the relaxing completeness problem.
Specifically, these two challenges are aligned with the challenges of form filling.
During data entry sessions, 
a form is usually partially filled and \approach needs to provide decisions on incomplete data.
Besides, in our context, we need to deal with large datasets since we
mainly target enterprise software systems that can collect a huge number of entries every day.

 \section{Discussion}
\label{sec:discussion}

\subsection{Usefulness}
\label{sec:usefulness}
The main goal of \approach is to prevent the entering of meaningless values
by relaxing the data entry form completeness requirements.
In order to assess the capability of \approach, 
we evaluated it with two real-world datasets,
including a public dataset from the 
biomedical domain and a proprietary dataset from the banking domain. 
These two datasets are related to existing data entry forms. 

Experiment results show that \approach outperforms baselines in
determining completeness requirements with a specificity score of at
least  {\LACQUERSeqSpecNcbi} and a NPV score higher than {\LACQUERSeqNPVProp}.
In the context of completeness requirement relaxation, these results mean that \approach can correctly (i.e., NPV $\ge$ {\LACQUERSeqNPVProp}) prevent the filling at least
20\% meaningless values.
In addition, \approach can correctly determine (with precision above {\LACQUERSeqPrec}) when a field should be required with a recall value of at least {\LACQUERRandRecProp}. 
This recall value means that \approach can almost determine all the required fields.
The high precision  value shows that \approach rarely incorrectly predicts optional fields as required.
In other words, \approach will not add much extra burden to users by adding more restrictions during the form filling process.

As discussed in section~\ref{sec:rq1}, \approach can determine more optional fields (i.e., a higher specificity)
in the PEIS dataset than in the NCBI dataset due to the higher data quality of the former.
Since we target data entry functionalities in enterprise software,
we expect to find similar conditions in other contexts in which data entry operators
follow corporate guidelines for selecting appropriate 
 values that should be filled when 
a field is not applicable. 
In such contexts, \approach is expected to provide results that are similar to
those achieved on the PEIS dataset. 

 \subsection{Practical Implications}
\label{sec:implication}

This subsection discusses the practical implications of \approach for
different stakeholders: software developers, end-users, and researchers.

\subsubsection{Software Developers}
\label{sec:implications:sw-des} 
\approach can help developers refactor data entry forms, which typically have many historical input instances and obsolete completeness requirements.
\approach does not require developers to define a complete set of rules 
regarding the completeness requirement of form fields. 
Developers can integrate \approach into a given data entry form as an independent tool. 
Deploying \approach into a data entry form requires providing a mapping between a data entry form, and field 
names and  column names in the dataset. 
The mapping needs only to be provided once and can be easily identified from Object Relational Mapping (ORM) and 
software design documentation. 
In addition to the mapping, deploying \approach requires a dictionary
of meaningless values, i.e., the values that should be used during the data entry process when a field is 
not applicable. 
We expect this dictionary to be found in the user manual of the data
entry software or in corporate guidelines, as it was the case for the PEIS dataset.

\subsubsection{End Users}
\label{sec:implications:end-users}
During the form filling process, obsolete required fields in the data entry form 
can affect the data accuracy since users have to enter meaningless values to 
skip filling these obsolete fields. 
 \approach  can automatically decide when a field should
be required or not based on the filled fields and historical input instances. 
Our experiments show that \approach can correctly determine between 20\% and 64\% of optional fields,
which reduces the  user effort and the time taken during the form filling process.

\subsubsection{Researchers}

In order to avoid predicting required field as optional, 
\approach includes an endorser module to decide if the prediction is accurate enough to be provided to the user.
We propose a novel strategy to automatically determine the threshold used in the endorser module.
Hence, our endorser module does not require any 
configuration from the domain expert.
We believe that such an endorser module can be
adopted by other researchers in other recommender systems.

 \subsection{Combining LACQUER with LAFF}
\label{sec:combiningLAFFandLACQUER}

\begin{figure}[tb]
	\centering{
\tikzstyle{field} = [rectangle, minimum width=1.2cm, minimum height=0.7cm, text centered, text width=2.3cm, draw=black, fill= white!30]

\tikzstyle{label} = [rectangle, minimum width=1.7cm, minimum height=0.35cm, text centered, text width=2.5cm, align=right, draw=white, fill= white!30]

\tikzstyle{title_node} = [rectangle, minimum width=0.5cm, minimum height=0.35cm, text centered, text width=4.5cm, draw=white, fill= white!30]

\tikzstyle{text_node} = [rectangle, minimum width=2.5cm, minimum height=0.7cm, text centered, text width=2.5cm, fill=white!30]

\tikzstyle{round_rect} = [rectangle, rounded corners, minimum width=1.5cm, minimum height=0.5 cm,text centered, draw=black, fill=white!30]

\newcommand{\DrawTriangle}[1][]{\begin{tikzpicture}[overlay,remember picture]
	\filldraw[fill=black] (90:1.2ex) -- (170:0.7ex) -- (10:0.7ex) --cycle;
	\end{tikzpicture}
}

\resizebox{1\textwidth}{!}{
\begin{tikzpicture}[node distance=3mm, >=latex]
\node (tb1) [field]{Wish};
	\node (f1) [label, left= of tb1] {Company Name};
	\node (tb2) [field, below=of tb1, text width=1.1cm, minimum width=1.1cm, xshift=-0.6cm] {};
	\node (f2) [label, left= of tb2] {Monthly revenue};
	\node (f2_unit) [label, right= of tb2, xshift=-2mm, align=left, text width=1.3cm] {\emph{k euro}};
	\node (tb3) [field, below=of tb2,xshift=0.6cm] {};
	\node (f3) [label, left= of tb3] {Company type};
	\node (tb4) [field, below=of tb3] {};
	\node (f4) [label, left= of tb4] {Field of activity};
	\node (tb5) [field, below=of tb4, text=red] {};
	\node (f5) [label, left= of tb5] {Tax ID};
	
	\node (icon_tb3) [isosceles triangle,
	isosceles triangle apex angle=60, draw, rotate=270, fill=gray!120, minimum size =0.1cm, right=of tb3, xshift=-0.25cm, yshift=-0.6cm]{};
	
	\node (icon_tb4) [isosceles triangle,
	isosceles triangle apex angle=60, draw, rotate=270, fill=gray!120, minimum size =0.1cm, right=of tb4, xshift=-0.25cm, yshift=-0.6cm]{};
		
	\node (submit) [round_rect, below= of tb5, xshift=0.35cm, minimum height=0.7cm, text centered, text width=1.6cm, fill=gray!30] {Submit};
	
	\node (cancel) [round_rect, left= of submit, xshift=-0.35cm, minimum height=0.7cm, text centered, text width=1.6cm, fill=gray!30] {Cancel};
	\node (lacquer_t1) [label, right= of f2_unit, text width= 2.5cm, xshift = -0.8 cm, minimum width =2.5cm ] {\scriptsize \approach :  Required};
	\node (ui) [draw=none, fit= (tb1) (tb2) (tb3)(tb4)(tb5)(f1)(f2)(f3)(f4)(f5)(submit) (lacquer_t1)] {};
	
	\node (title)[title_node, font=\fontsize{12}{0}\selectfont, above=of ui] {\textbf{Data entry form \emph{F}}};
	
	\node (form) [draw=black, fit= (ui)(title), minimum width = 8cm] {};

	\node (tb1_t2) [field, right= of f1, xshift =11cm]{Wish};
	\node (f1_t2) [label, left= of tb1_t2] {Company Name};
	\node (tb2_t2) [field, below=of tb1_t2, text width=1.1cm, minimum width=1.1cm, xshift=-0.6cm] {20};
	\node (f2_t2) [label, left= of tb2_t2] {Monthly revenue};
	\node (f2_unit_t2) [label, right= of tb2_t2, xshift=-2mm, align=left, text width=1.3cm] {\emph{k euro}};
	\node (tb3_t2) [field, below=of tb2_t2,xshift=0.6cm] {};
	\node (f3_t2) [label, left= of tb3_t2] {Company type};
	\node (tb4_t2) [field, below=of tb3_t2] {};
	\node (f4_t2) [label, left= of tb4_t2] {Field of activity};
	\node (tb5_t2) [field, below=of tb4_t2] {};
	\node (f5_t2) [label, left= of tb5_t2] {Tax ID};
	
	\node (icon_tb3_t2) [isosceles triangle,
	isosceles triangle apex angle=60, draw, rotate=270, fill=gray!120, minimum size =0.1cm, right=of tb3_t2, xshift=-0.25cm, yshift=-0.6cm]{};
	
	\node (icon_tb4_t2) [isosceles triangle,
	isosceles triangle apex angle=60, draw, rotate=270, fill=gray!120, minimum size =0.1cm, right=of tb4_t2, xshift=-0.25cm, yshift=-0.6cm]{};
	
	\node (submit_t2) [round_rect, below= of tb5_t2, xshift=0.35cm, minimum height=0.7cm, text centered, text width=1.6cm, fill=gray!30] {Submit};
	
	\node (cancel_t2) [round_rect, left= of submit_t2, xshift=-0.35cm, minimum height=0.7cm, text centered, text width=1.6cm, fill=gray!30] {Cancel};
	
	\node (lacquer_t2) [label, right= of tb3_t2, text width= 2.5cm, xshift = -0.3 cm, minimum width =2.5cm,yshift= 0.2cm ] {\scriptsize \approach :  Required};
	\node (laff_t2) [label, below = of lacquer_t2, text width= 1.5cm, minimum width =1.5cm , yshift= 0.3cm , xshift = -0.45 cm] {\scriptsize LAFF:  NPO};
	
	\node (ui_t2) [draw=none, fit= (tb1_t2) (tb2_t2) (tb3_t2)(tb4_t2)(tb5_t2)(f1_t2)(f2_t2)(f3_t2)(f4_t2)(f5_t2)(submit_t2) (lacquer_t2) , right = of ui ] {};
	
	\node (title_t2)[title_node, font=\fontsize{12}{0}\selectfont, above=of ui_t2, xshift= 0.75cm] {\textbf{Data entry form \emph{F}}};
	
	\node (form_t2) [draw=black, fit= (ui_t2)(title_t2),minimum width = 8cm, xshift= 2cm] {};

	\node (tb1_t3) [field, below= of form,yshift=-1.2cm]{Wish};
	\node (f1_t3) [label, left= of tb1_t3] {Company Name};
	\node (tb2_t3) [field, below=of tb1_t3, text width=1.1cm, minimum width=1.1cm, xshift=-0.6cm] {20};
	\node (f2_t3) [label, left= of tb2_t3] {Monthly revenue};
	\node (f2_unit_t3) [label, right= of tb2_t3, xshift=-2mm, align=left, text width=1.3cm] {\emph{k euro}};
	\node (tb3_t3) [field, below=of tb2_t3,xshift=0.6cm] {NPO};
	\node (f3_t3) [label, left= of tb3_t3] {Company type};
	\node (tb4_t3) [field, below=of tb3_t3] {};
	\node (f4_t3) [label, left= of tb4_t3] {Field of activity};
	\node (tb5_t3) [field, below=of tb4_t3] {};
	\node (f5_t3) [label, left= of tb5_t3] {Tax ID};
	
	\node (icon_tb3_t3) [isosceles triangle,
	isosceles triangle apex angle=60, draw, rotate=270, fill=gray!120, minimum size =0.1cm, right=of tb3_t3, xshift=-0.25cm, yshift=-0.6cm]{};
	
	\node (icon_tb4_t3) [isosceles triangle,
	isosceles triangle apex angle=60, draw, rotate=270, fill=gray!120, minimum size =0.1cm, right=of tb4_t3, xshift=-0.25cm, yshift=-0.6cm]{};
	
	\node (submit_t3) [round_rect, below= of tb5_t3, xshift=0.35cm, minimum height=0.7cm, text centered, text width=1.6cm, fill=gray!30] {Submit};
	
	\node (cancel_t3) [round_rect, left= of submit_t3, xshift=-0.35cm, minimum height=0.7cm, text centered, text width=1.6cm, fill=gray!30] {Cancel};
	
	\node (lacquer_t3) [label, right= of tb4_t3, text width= 2.5cm, xshift = -0.3 cm, minimum width =2.5cm,yshift= 0.2cm ] {\scriptsize \approach :  Required};
	\node (laff_t3) [label, below = of lacquer_t3, text width= 1.9cm, minimum width =2.5cm , yshift= 0.35cm , xshift = -0.1 cm] {\scriptsize LAFF:  Education};
	
	\node (ui_t3) [draw=none, fit= (tb1_t3) (tb2_t3) (tb3_t3)(tb4_t3)(tb5_t3)(f1_t3)(f2_t3)(f3_t3)(f4_t3)(f5_t3)(submit_t3) (lacquer_t3) , below = of ui ,yshift = -1.2cm] {};
	
	\node (title_t3)[title_node, font=\fontsize{12}{0}\selectfont, above=of ui_t3, xshift= 0.75cm] {\textbf{Data entry form \emph{F}}};
	
	\node (form_t3) [draw=black, fit= (ui_t3)(title_t3),minimum width = 8.2cm] {};

	\node (tb1_t4) [field, below= of form_t2,yshift=-1.2cm]{Wish};
	\node (f1_t4) [label, left= of tb1_t4] {Company Name};
	\node (tb2_t4) [field, below=of tb1_t4, text width=1.1cm, minimum width=1.1cm, xshift=-0.6cm] {20};
	\node (f2_t4) [label, left= of tb2_t4] {Monthly revenue};
	\node (f2_unit_t4) [label, right= of tb2_t4, xshift=-2mm, align=left, text width=1.3cm] {\emph{k euro}};
	\node (tb3_t4) [field, below=of tb2_t4,xshift=0.6cm] {NPO};
	\node (f3_t4) [label, left= of tb3_t4] {Company type};
	\node (tb4_t4) [field, below=of tb3_t4] {Education};
	\node (f4_t4) [label, left= of tb4_t4] {Field of activity};
	\node (tb5_t4) [field, below=of tb4_t4] {};
	\node (f5_t4) [label, left= of tb5_t4] {Tax ID};
	
	\node (icon_tb3_t4) [isosceles triangle,
	isosceles triangle apex angle=60, draw, rotate=270, fill=gray!120, minimum size =0.1cm, right=of tb3_t4, xshift=-0.25cm, yshift=-0.6cm]{};
	
	\node (icon_tb4_t4) [isosceles triangle,
	isosceles triangle apex angle=60, draw, rotate=270, fill=gray!120, minimum size =0.1cm, right=of tb4_t3, xshift=-0.25cm, yshift=-0.6cm]{};
	
	\node (submit_t4) [round_rect, below= of tb5_t4, xshift=0.35cm, minimum height=0.7cm, text centered, text width=1.6cm, fill=gray!30] {Submit};
	
	\node (cancel_t4) [round_rect, left= of submit_t4, xshift=-0.35cm, minimum height=0.7cm, text centered, text width=1.6cm, fill=gray!30] {Cancel};
	
	\node (lacquer_t4) [label, right= of tb5_t4, text width= 2.5 cm, xshift = -0.3 cm, minimum width =2.5cm,yshift= 0.2cm ] {\scriptsize \approach :  N/A};
	\node (laff_t4) [label, below = of lacquer_t4, text width= 1.9cm, minimum width =2.5cm , yshift= 0.35cm , xshift = -0.1 cm] {\scriptsize LAFF:  N/A};
	
	\node (ui_t4) [draw=none, fit= (tb1_t4) (tb2_t4) (tb3_t4)(tb4_t4)(tb5_t4)(f1_t4)(f2_t4)(f3_t4)(f4_t4)(f5_t4)(submit_t4)(lacquer_t4)(laff_t4) , below = of ui_t2 ,yshift = -1.2cm] {};
	
	\node (title_t4)[title_node, font=\fontsize{12}{0}\selectfont, above=of ui_t4, xshift= 0.75cm] {\textbf{Data entry form \emph{F}}};
	
	\node (form_t4) [draw=black, fit= (ui_t4)(title_t4),minimum width = 8.2cm,xshift= 2cm] {};

	\node[shape=circle, draw, fill=gray, opacity=.2, text opacity=1, inner sep=0.5pt, right =of form, xshift=-1cm, yshift=3cm] (1) {1};
	\node[shape=circle, draw, fill=gray, opacity=.2, text opacity=1, inner sep=0.5pt, right =of form_t2, xshift=-1cm, yshift=3cm] (2) {2};
	\node[shape=circle, draw, fill=gray, opacity=.2, text opacity=1, inner sep=0.5pt, right =of form_t3, xshift=-1cm, yshift=3cm] (3) {3};
	\node[shape=circle, draw, fill=gray, opacity=.2, text opacity=1, inner sep=0.5pt, right =of form_t4, xshift=-1cm, yshift=3cm] (4) {4};

\end{tikzpicture}
}
 }
	\caption{Use case to combine \approach and LAFF together during form filling}
	
	\label{fig:use-case}
\end{figure}

Despite the differences explained in section~\ref{sec:related_work}, 
\approach and LAFF are complementary in practice. 
Both approaches can be combined as an AI-based assistant for form filling
to help users fill forms and ensure better data quality.

\figurename~\ref{fig:use-case} shows a possible scenario that uses both approaches together during a form-filling session. 
In this example, we assume that the user follows the sequential filling order. 
First, after filling in the company name field, 
\approach can already  check whether the ``monthly income'' field is required or not. 
Since ``monthly income'' is a numerical field, LAFF cannot perform a
prediction (LAFF only supports categorial fields). 
In this example, \approach determines that the field is required,
hence the user should fill it out. 
The ``Company type'' and ``Field of activity'' fields are both categorical. 
For these two fields, based on the filled fields, first \approach
determines the completeness requirement for each field.
Once the user clicks on a field, LAFF is enabled to provide a ranked list of possible values that can be used for this field. 
If the decision of \approach on a field is optional,
LAFF can still be activated to provide suggestions as long as the user wants to fill in the field. 
Finally, let us assume that the ``Tax ID'' field (a numerical one) is optional by
design. In this case, both LAFF and \approach are not enabled, since
there is no need for \approach to relax a
completeness requirement and the field is numerical and thus not compatible with
LAFF.

 \section{Conclusion}
\label{sec:conclusion}

In this paper we proposed \approach, an approach to automatically relax the completeness 
requirement of data entry forms by deciding when a field should be optional 
based on the filled fields and historical input instances. 
\approach applies Bayesian Networks on an oversampled data set (using SMOTE) to learn 
the completeness requirement dependencies between fields. 
Moreover, \approach uses a heuristic-based endorser module to ensure that it only 
provides accurate suggestions. 

We  evaluated \approach on two datasets, one proprietary dataset from the banking domain 
and one public dataset from the biomedical domain. 
Our results show that \approach can correctly determine 20\% to 64\% of optional fields
and determine almost all the required fields (with a recall value of \LACQUERRandRecProp).
\approach takes at most {\SI{\LACQUERPredictMax}{\milli\s}}  to provide a suggestion,
which complies with human-computer interaction principles to ensure a seamless interaction with users. 

As a part of future work, we plan to conduct a user study to analyze the effect of \approach 
in reducing the meaningless values and the effort spent by  users during the form filling
process. 
We plan also to add an automated module that can detect meaningless values 
entered by the users during form filling, when such values have 
not been specified by the form designer.
Furthermore, we plan to integrate \approach into platforms for the design of data entry forms~\cite{gravityforms, surveymonkey, googleforms}
to help designers perform form refactoring. 
These platforms currently rely on rules defined by designers to specify completeness requirements during the design phase. 
\approach can be used to relieve designers from the task of defining
such rules, since it only requires to indicate the required fields;
during form filling, \approach will automatically suggest the
completeness requirement of the required fields. 
\approach can also be extended to support sophisticated input fields
that can handle multiple selections such us drop-down menus and checkbox groups.
Finally, we plan to extend \approach to support updates of existing
data entries as well as to determine whether fields previously marked
as optional should become required.

\subsection*{Acknowledgements}
\label{sec:acknowledgements}
Financial support for this work was provided by the Alphonse Weicker
Foundation and by our industrial partner BGL BNP Paribas
Luxembourg. We thank Anne Goujon, Michael Stanisiere, and Fernand
Lepage for their help with the PEIS dataset; we thank  Clément Lefebvre Renard and Andrey Boytsov for their comments on earlier drafts of the paper.

 \bibliographystyle{ACM-Reference-Format}

\end{document}